\documentclass{article}
\PassOptionsToPackage{numbers}{natbib}

\usepackage[final]{neurips_2020}
\usepackage[utf8]{inputenc}                 % allow utf-8 input
\usepackage[T1]{fontenc}                    % use 8-bit T1 fonts
\usepackage{xr-hyper}
\usepackage[colorlinks]{hyperref}           % hyperlinks
\usepackage{url}                            % simple URL typesetting
\usepackage{booktabs}                       % professional-quality tables
\usepackage{amsfonts}                       % blackboard math symbols
\usepackage{nicefrac}                       % compact symbols for 1/2, etc.
\usepackage{microtype}                      % microtypography

\usepackage{bm}                             % bold math symbols
\usepackage{algorithm}                      % environment algorithm 
\usepackage{algpseudocode}                  % algorithm formatting
\usepackage{setspace}                       % pleasant display of algorithm.
\usepackage{amsmath}                        % equation*
\usepackage{subcaption}                     % table with caption
\usepackage{multirow}                       % complex formatting of cells
\usepackage{graphicx}                       % scalebox
\usepackage{xcolor}                         % textcolor
\usepackage{enumitem}
\usepackage{mathtools}                      % \coloneqq
\usepackage{calc}                           %\widthof{}
\usepackage{xspace}
\usepackage{array}
\usepackage[resetlabels]{multibib}
\usepackage{footmisc}

\usepackage{silence}
\definecolor{BrickRed}{rgb}{0.6,0,0}
\definecolor{RoyalBlue}{rgb}{0,0,0.8}
\hypersetup{citecolor=BrickRed, linkcolor=RoyalBlue, filecolor=RoyalBlue}
\newcommand{\algname}{genetic expert-guided learning\xspace}
\newcommand{\Algname}{Genetic expert-guided learning\xspace}
\newcommand{\ALGname}{GEGL\xspace}

\newcommand{\ph}[1]{\phantom{#1}}

\newcommand{\authoryear}[1]{[\citeauthor{#1}~\citeyear{#1}]}
\newcommand{\PreserveBackslash}[1]{\let\temp=\\#1\let\\=\temp}
\newcolumntype{C}[1]{>{\PreserveBackslash\centering}p{#1}}
\newcolumntype{R}[1]{>{\PreserveBackslash\raggedleft}p{#1}}
\newcolumntype{L}[1]{>{\PreserveBackslash\raggedright}p{#1}}

\DeclareMathOperator{\argmin}{arg\,min} % thin space, limits on side in displays

\def\piex{\ensuremath\pi_{\mathtt{ex}}}
\def\queue{\ensuremath\mathcal{Q}}
\def\queueex{\ensuremath\mathcal{Q}_{\mathtt{ex}}}
\def\plogp{$\mathtt{PenalizedLogP}$\xspace}
\newcommand{\stdv}[1]{\scriptsize~$\pm$~#1}

\algnewcommand{\LeftComment}[1]{$\qquad$\(\triangleright\) \textit{#1}}
\title{Guiding Deep Molecular Optimization \\ with Genetic Exploration}

\author{
  Sungsoo Ahn\hspace{.2in}Junsu Kim\hspace{.2in}Hankook Lee\hspace{.2in}Jinwoo Shin\\
  Korea Advanced Institute of Science and Technology (KAIST)\\
  \texttt{\{sungsoo.ahn, junsu.kim, hankook.lee, jinwoos\}@kaist.ac.kr}
}

\begin{document}

\maketitle

\begin{abstract}
De novo molecular design attempts to search over the chemical space for molecules with the desired property. Recently, deep learning has gained considerable attention as a promising approach to solve the problem. In this paper, we propose \algname (\ALGname), a simple yet novel framework for training a deep neural network (DNN) to generate highly-rewarding molecules. Our main idea is to design a ``genetic expert improvement'' procedure, which generates high-quality targets for imitation learning of the DNN. Extensive experiments show that \ALGname significantly improves over state-of-the-art methods. For example, \ALGname manages to solve the penalized octanol-water partition coefficient optimization with a score of $31.40$, while the best-known score in the literature is $27.22$. Besides, for the GuacaMol benchmark with 20 tasks, our method achieves the highest score for 19 tasks, in comparison with state-of-the-art methods, and newly obtains the perfect score for three tasks. Our training code is available at \url{https://github.com/sungsoo-ahn/genetic-expert-guided-learning}.
\end{abstract}

\section{Introduction}
Discovering new molecules with the desired property is fundamental in chemistry, with critical applications such as drug discovery \citep{hughes2011principles} and material design \citep{yan2018non}. The task is challenging since the molecular space is vast; e.g., the number of synthesizable drug-like compounds is estimated to be around $10^{60}$ \citep{bohacek1996art}. To tackle this problem, \textit{de novo molecular design} \citep{schneider2005computer, schneider2013novo} aims to generate a new molecule from scratch with the desired property, rather than na\"ively enumerate over the molecular space.

Over the past few years, molecule-generating deep neural networks (DNNs) have demonstrated successful results for solving the de novo molecular design problem \citep{guimaraes2017objective, kusner2017grammar, gomez2018automatic, you2018graph, liu2018constrained, dai2018syntaxdirected, jin2018junction, neil2018exploring, assouel2018defactor, bradshaw2019model, popova2019molecularrnn, winter2019efficient, zhou2019optimization, shi2020graphaf,  griffiths2020constrained, jin2020hierarchical, gottipati2020learning}. 
For example, \citet{gomez2018automatic} perform Bayesian optimization for maximizing the desired property, on the embedding space of molecule-generating variational auto-encoders.
%For example, \citet{gomez2018automatic} performs Bayesian optimization on the embedding space of molecule-generating variational auto-encoders for maximization of the desired property, respectively.
On the other hand, \citet{guimaraes2017objective} train a molecule-generating policy using reinforcement learning with the desired property formulated as a reward.

Intriguingly, several works \citep{yoshikawa2018population, jensen2019graph, nigam2020augmenting, polishchuk2020crem} have recently evidenced that the traditional frameworks based on genetic algorithm (GA) can compete with or even outperform the recently proposed deep learning methods. They reveal that GA is effective, thanks to the powerful domain-specific genetic operators for exploring the chemical space. For example, \citet{jensen2019graph} achieves outstanding performance by generating new molecules as a combination of subgraphs extracted from existing ones. Such observations also emphasize how domain knowledge can play a significant role in de novo molecular design. On the contrary, the current DNN-based methods do not exploit such domain knowledge explicitly; instead, they implicitly generalize the knowledge of high-rewarding molecules by training a DNN on them. Notably, the expressive power of DNN allows itself to parameterize a distribution over the whole molecular space flexibly.

\begin{figure}
    \centering
    \includegraphics[width=0.9\textwidth]{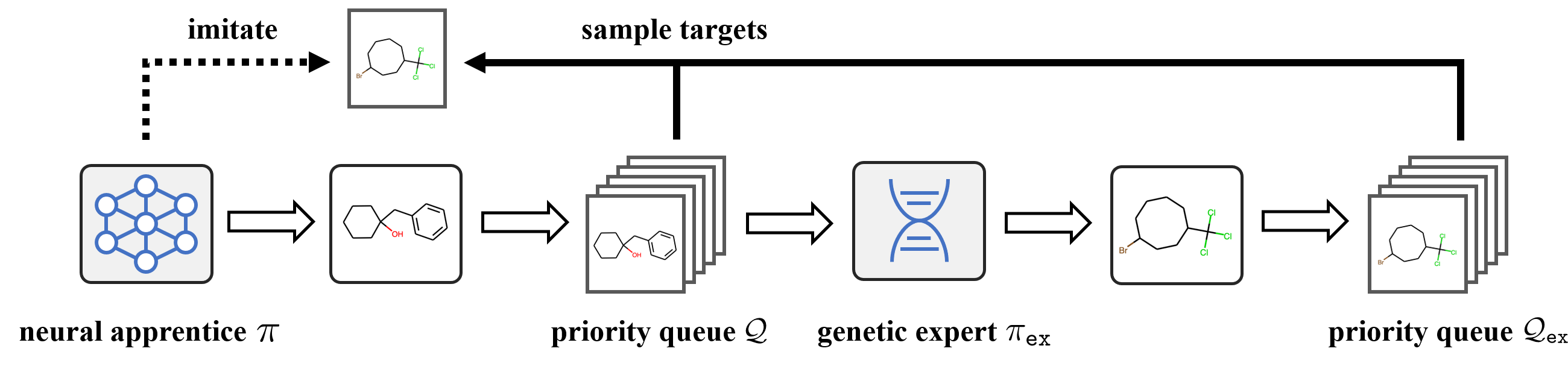}
    \caption{Illustration of the proposed \algname (\ALGname) framework.}
    \label{fig:concept}
    \vspace{-.1in}
\end{figure}

\textbf{Contribution.} 
%\magenta{In this work, we propose \algname (\ALGname), which is a novel framework; it aims at training a molecule-generating DNN with additional genetic exploration.}
In this work, we propose \algname (\ALGname), which is a novel framework for training a molecule-generating DNN guided with genetic exploration. Our main idea is to formulate an \textit{expert policy} by applying the domain-specific genetic operators, i.e., mutation and crossover,\footnote{See Figure \ref{fig:gen_operators} for illustrations of mutation and crossover.} to the DNN-generated molecules. Then the DNN becomes an \textit{apprentice policy} that learns to imitate the highly-rewarding molecules discovered by the expert policy. Since the expert policy improves over the apprentice policy by design, the former policy consistently guides the latter policy to generate highly-rewarding molecules. We provide an overall illustration of our framework in Figure \ref{fig:concept}. 

%\magenta{The high-level idea is to use the genetic operators, i.e., mutation and crossover,\footnote{See Figure \ref{fig:gen_operators} for illustrations of mutation and crossover.} to formulate an ``expert improvement operator''. The operator allows the DNN to imitate an improved version of itself.}
%  

%The high-level idea is to use the genetic operators, i.e., mutation and crossover,\footnote{See Figure \ref{fig:gen_operators} for illustrations of mutation and crossover.} to formulate an ``expert improvement operator'', which allows the DNN to imitate an improved version of itself.
%\textcolor{red}{Namely, we formulate the \textit{expert policy} by applying the genetic operators to the molecules obtained from the DNN.} Then the DNN becomes an \textit{apprentice policy} that learns to generalize the knowledge of highly-rewarding molecules discovered by the expert policy. We provide an overall illustration of our framework in Figure \ref{fig:concept}. 

%Since the expert policy is additionally equipped with genetic operators for exploration, it can generate molecules with higher quality than that of the DNN. 

We note that our \ALGname framework can be seen as a reinforcement learning algorithm with a novel mechanism for additional explorations. To be specific, the generation of a molecule can be regarded as an action and the desired property of the generated molecule as a reward. Similar to most reinforcement learning algorithms, reducing the sample complexity is crucial in our \ALGname framework. To this end, we design our framework with \textit{max-reward priority queues} \citep{neil2018exploring, gupta2019feedback, liang2017neural, abolafia2018neural,agarwal2019learning}. By storing the highly-rewarding molecules, the priority queues prevent the policies from ``forgetting'' the valuable knowledge. 

We extensively evaluate our method on four experiments: (a) optimization of penalized octanol-water partition coefficient, (b) optimization of penalized octanol-water partition coefficient under similarity constraints, (c) the GuacaMol benchmark \citep{brown2019guacamol} consisting of 20 de novo molecular design tasks, and (d) the GuacaMol benchmark evaluated under post-hoc filtering procedure \citep{gao2020synthesizability}. Remarkably, our \ALGname framework outperforms all prior methods for de novo molecular design by a large margin. In particular, \ALGname achieves the penalized octanol-water partition coefficient score of $31.40$, while the best baseline \citep{gottipati2020learning} and the second-best baseline \citep{winter2019efficient} achieves the score of $27.22$ and $26.1$, respectively. For the GuacaMol benchmark, our algorithm achieves the highest score for 19 out of 20 tasks and newly achieves the perfect score for three tasks.

\section{Related works}
Automating the discovery of new molecules is likely to have significant impacts on essential applications such as drug discovery \citep{hughes2011principles} and material design \citep{yan2018non}. To achieve this, researchers have traditionally relied on \textit{virtual screening} \citep{shoichet2004virtual}, which typically works in two steps: (a) enumerating all the possible combinations of predefined building-block molecules and (b) reporting the molecules with the desired property. However, the molecular space is large, and it is computationally prohibitive to enumerate and score the desired property for all the possible molecules. 

\textit{De novo molecular design} \citep{schneider2005computer} methods attempt to circumvent this issue by generating a molecule from scratch. Instead of enumerating a large set of molecules, these methods search over the molecular space to maximize the desired property.
%In this way, the task of de novo molecular design solves the problem of molecular discovery as a combinatorial optimization over the molecular space for maximizing the desired property. 
%Given the promise of de novo molecular design, researchers have developed a considerable amount of studies for this task. 
In the following, we discuss the existing methods for de novo molecular design categorized by their optimization schemes: deep reinforcement learning, deep embedding optimization, and genetic algorithm. %We also provide more detailed descriptions of the works using deep learning to model molecules in Appendix \ref{sec:detailed_related}.

\textbf{Deep reinforcement learning (DRL).} 
Deep reinforcement learning is arguably the most straightforward approach for solving the de novo molecular design problem with deep neural networks (DNNs) \citep{guimaraes2017objective, you2018graph, neil2018exploring, popova2019molecularrnn, zhou2019optimization, shi2020graphaf, gottipati2020learning}. Such methods formulate the generation of the molecules as a Markov decision process. In such formulations, the desired properties of molecules correspond to high rewards, and the policy learns to generate highly-rewarding molecules. We also note several works using deep reinforcement learning to solve combinatorial problems similar to the task of de novo molecular design, e.g., program synthesis \citep{liang2017neural}, combinatorial mathematics \citep{bello2016neural}, biological sequence design \citep{angermueller2020model}, and protein structure design \citep{brookes2019conditioning}. 

\textbf{Deep embedding optimization (DEO).}
Also, there exist approaches to optimize molecular ``embeddings'' extracted from DNNs \citep{kusner2017grammar, gomez2018automatic, liu2018constrained, dai2018syntaxdirected, jin2018junction, bradshaw2019model, winter2019efficient, griffiths2020constrained}. Such methods are based on training a neural network to learn a mapping between a continuous embedding space and a discrete molecular space. Then they apply optimization over the continuous embedding to maximize the desired property of the corresponding molecule. Methods such as gradient descent \citep{liu2018constrained, bradshaw2019model}, Bayesian optimization \citep{kusner2017grammar, gomez2018automatic, dai2018syntaxdirected, jin2018junction}, constrained Bayesian optimization \citep{griffiths2020constrained}, and particle swarm optimization \citep{winter2019efficient} have been applied for continuous optimization of the embedding space. 

\textbf{Genetic algorithm (GA).} 
Inspired from the concept of natural selection, genetic algorithms (GAs) \citep{yoshikawa2018population, jensen2019graph, polishchuk2020crem, weininger1995method, globus1999automatic, douguet2000genetic, schneider2000novo, brown2004graph} search over the molecular space with genetic operators, i.e., mutation and crossover. To this end, the mutation randomly modifies an existing molecule, and the crossover randomly combines a pair of molecules. Such approaches are appealing since they are simple and capable of incorporating the domain knowledge provided by human experts. While the deep learning methods are gaining attention for de novo molecular design, several works \citep{yoshikawa2018population, jensen2019graph, polishchuk2020crem} have recently demonstrated that GA can compete with or even outperform the DNN-based methods.

%Reinforcement learning is a popular approach for solving combinatorial tasks with deep neural networks Another approach for de novo molecular design is based on reinforcement learning \citep{xx}. Typically, networks are pre-trained on a dataset \citep{xx} to ``transfer'' the knowledge of molecules. Notably, \citet{xx} introduced a discriminator-based reward term to regularize the generated molecules to be more realistic. On the other hand, \citet{xx} starts the training from scratch to perform an unbiased search for novel molecules. Apart from de novo molecular desgin tasks, deep reinforcement learning has also been applied to solving combinatorial optimization problems in combinatorial mathematics \citep{bello2016neural}, program synthesis \citep{liang2017neural}, biological sequence design \citep{angermueller2020model}, and protein structure design \citep{brookes2019conditioning}. 

%Apart from de novo molecular design tasks, genetic algorithms are popular for solving problems including, but not limited to, the traveling salesman problem \citep{xx}, finding hardware bugs \citep{xx}, ruleset production \citep{xx}, climatology \citep{xx}, and economics \citep{xx}. 

%\newpage
\section{\Algname (\ALGname)}
\subsection{Overview of \ALGname} 

In this section, we introduce \algname (\ALGname), a novel yet simple framework for de novo molecular design. To discover highly-rewarding molecules, GEGL aims at training a molecule-generating deep neural network (DNN). Especially, we design the framework using an additional genetic \textit{expert policy}, which generates targets for imitation learning of the neural \textit{apprentice policy}, i.e., the DNN. Our main idea is about formulating the expert policy as a ``genetic improvement operator'' applied to the apprentice policy; this allows us to steer the apprentice policy towards generating highly-rewarding molecules by imitating the better expert policy. 

To apply our framework, we view de novo molecular design as a combinatorial optimization of discovering a molecule $\bm{x}$, which maximizes the \textit{reward} $r(\bm{x})$, i.e., the desired property.\footnote{In this paper, we consider the properties of molecules which can be measured quantitatively.} To solve this problem, we collect highly-rewarding molecules from the neural apprentice policy $\pi(\bm{x}; \theta)$ and the genetic expert policy $\piex(\bm{x}; \mathcal{X})$ throughout the learning process. Here, $\theta$ indicates the DNN's parameter representing the apprentice policy, and $\mathcal{X}$ indicates a set of ``seed'' molecules to apply the genetic operators for the expert policy. Finally, we introduce fixed-size \textit{max-reward priority queues} $\queue$ and $\queueex$, which are buffers that only keep a fixed number of molecules with the highest rewards.

Our \ALGname framework repeats the following three-step procedure: 
\begin{itemize}[leftmargin=0.6in]
\item[\textbf{Step A.}] The apprentice policy $\pi(\bm{x}; \theta)$ generates a set of molecules. Then the max-reward priority queue $\queue$ with the size of $K$ stores the generated molecules.

\item[\textbf{Step B.}] The expert policy $\piex(\bm{x}; \queue)$ generates molecules using the updated priority queue $\queue$ as the seed molecules. Next, the priority queue $\queueex$ with a size of $K$ stores the generated molecules.

\item[\textbf{Step C.}] The apprentice policy optimizes its parameter $\theta$ by learning to imitate the molecules sampled from the union of the priority queues $\queue \cup \queueex$. In other words, the parameter $\theta$ is updated to maximize $\sum_{\bm{x}\in \queue \cup \queueex}\log\pi(\bm{x};\theta)$.
%In other words, the apprentice policy learns to maximize the log-likelihood of the molecules in $\queue \cup \queueex$. 
\end{itemize}

\begin{figure}[t]
\centering
\begin{minipage}[t!]{0.9\textwidth}
\centering
\begin{algorithm}[H]
\small
\begin{spacing}{1.2}
\caption{\Algname (\ALGname)}
\label{alg:genetic_exploration_guided_learning}
\small
\begin{algorithmic}[1]
\State 
Set $\queue\gets \emptyset$, $\queueex\gets \emptyset$. \Comment{{\it Initialize the max-reward priority queues $\queue$ and $\queueex$.}}
\For{$t=1,\ldots, T$}
    \For{$m=1,\ldots, M$} 
    \Comment{{\it Step A: add $M$ samples generated by $\pi$ into $\queue$.}}
    \State Update $\queue \leftarrow \queue \cup \{\bm{x}\}$, where $\bm{x} \sim \pi(\bm{x};\theta)$.
    \State If $|\queue|>K$, update $\queue \leftarrow \queue\setminus\{\bm{x}_\mathtt{min}\}$, where $\bm{x}_\mathtt{min} = \argmin_{\bm{x} \in \queue} r(\bm{x})$.
    \EndFor
    \For{$m=1,\ldots, M$} 
    \Comment{{\it Step B: add $M$ samples generated by $\piex$ into $\queueex$.}}
    \State Update $\queueex \leftarrow \queueex \cup \{\bm{x}\}$, where $\bm{x} \sim \piex(\bm{x}; \queue)$.
    \State If $|\queueex| > K$, update $\queueex \leftarrow \queueex \setminus \{\bm{x}_\mathtt{min}\},$ where $\bm{x}_\mathtt{min} = \argmin_{\bm{x} \in \queueex} r(\bm{x})$.
    \EndFor
    \State Maximize $\sum_{\bm{x}\in \queue \cup \queueex}\log \pi(\bm{x};\theta)$ over $\theta$.
    \Comment{{\it Step C: train $\pi$ with imitation learning.}}
\EndFor
\State Report $\queue \cup \queueex$ as the output. 
\Comment{{\it Output the highly-rewarding molecules.}}
\end{algorithmic}
\end{spacing}
\end{algorithm}
\end{minipage}
\vspace{-.2in}
\end{figure}

One may interpret GEGL as a deep reinforcement learning algorithm. From such a perspective, the corresponding Markov decision process has a fixed episode-length of one, and its action corresponds to the generation of a molecule. Furthermore, \ALGname highly resembles the prior works on expert iteration \citep{anthony2017thinking, anthony2019policy} and AlphaGo Zero \citep{silver2017mastering}, where the Monte Carlo tree search is used as an improvement operator that guides training through enhanced exploration. We provide an illustration and a detailed description of our algorithm in Figure \ref{fig:concept} and Algorithm \ref{alg:genetic_exploration_guided_learning}, respectively.

\subsection{Detailed components of \ALGname}
In the rest of this section, we provide a detailed description of each component in \ALGname. We start by explaining our design choices on the expert and the apprentice policies. Then we discuss how the max-reward priority queues play an essential role in our framework.

\textbf{Genetic expert policy.}
Our genetic expert policy $\piex(x;\mathcal{X})$ is a distribution induced by applying the genetic operators, i.e., mutation and crossover, to a set of molecules $\mathcal{X}$. We use the genetic operators highly optimized (with domain knowledge) for searching over the molecular space; hence the expert policy efficiently improves over the apprentice policy in terms of exploration. 

\begin{figure}[t]
    \centering
    \begin{subfigure}{.47\textwidth}
    \centering
    \hspace*{\fill}
    \includegraphics[width=1.0\textwidth]{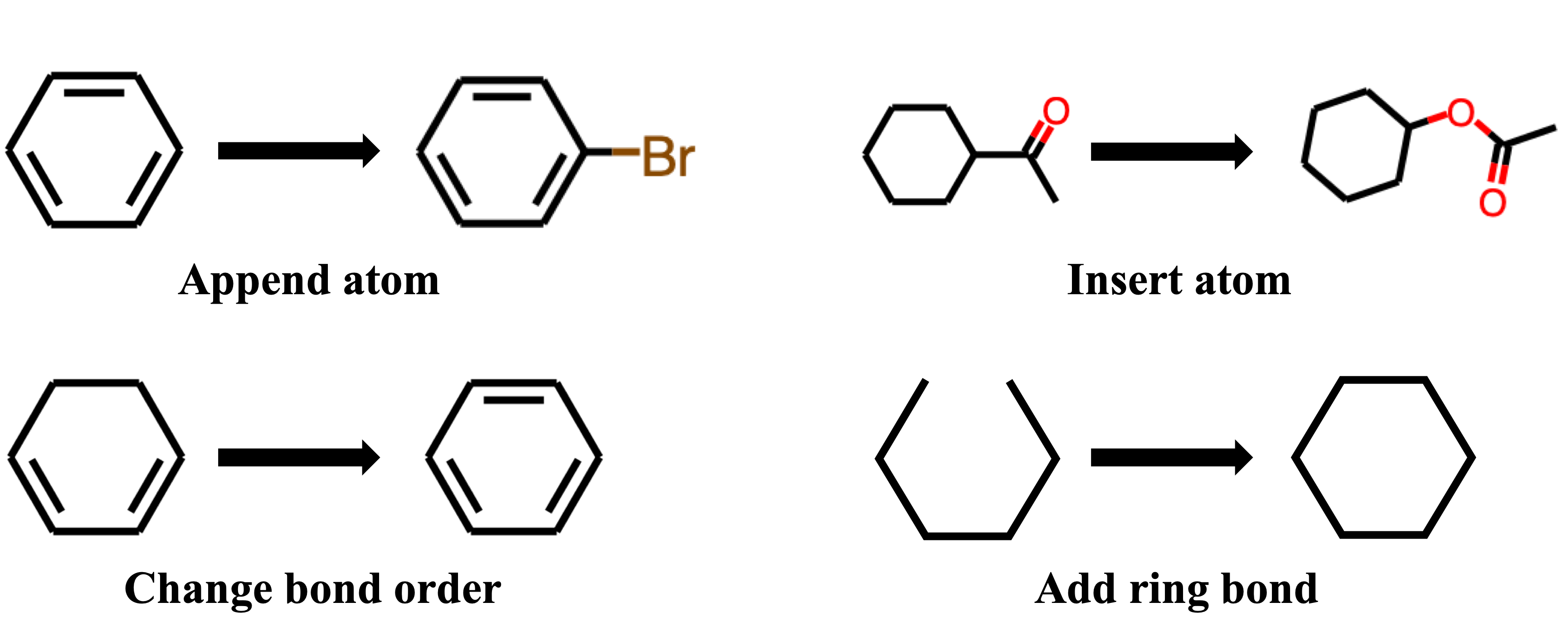}
    \caption{Mutation}
    \label{subfig:mutation}
    \end{subfigure}
    \hspace*{\fill}
    \begin{subfigure}{.47\textwidth}
    \centering
    \includegraphics[width=1.0\textwidth]{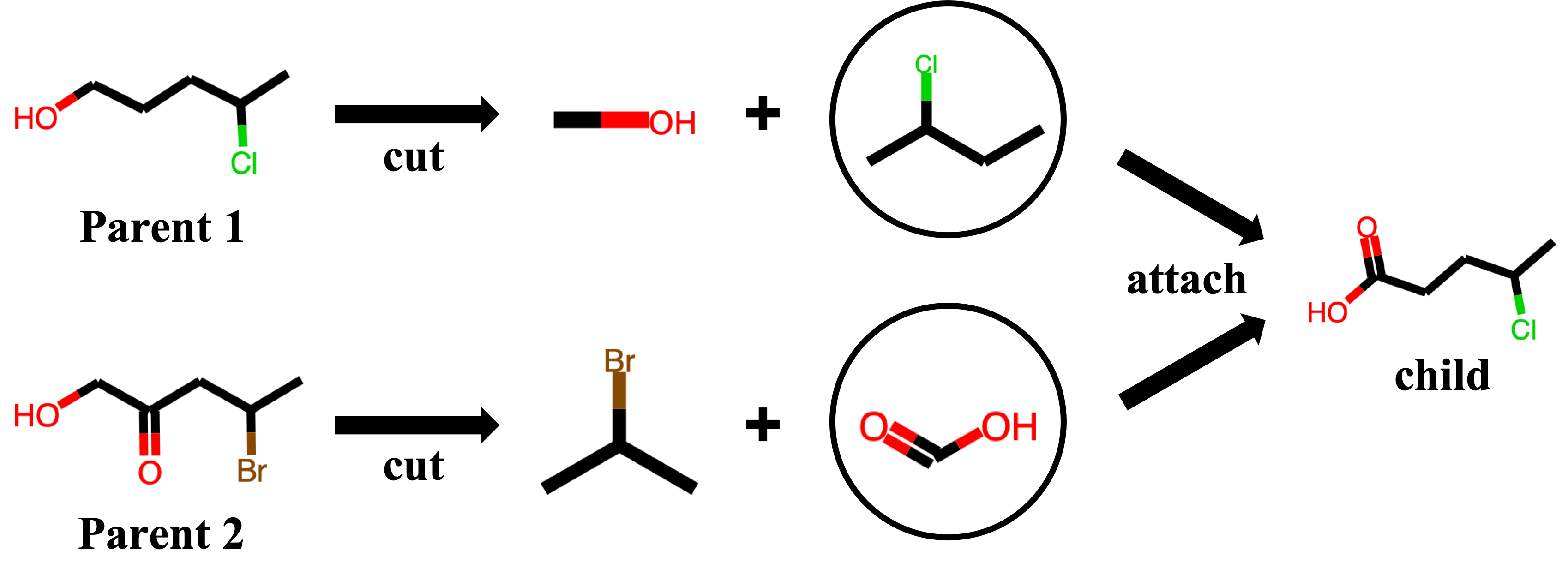}
    \caption{Crossover}
    \label{subfig:crossover}
    \end{subfigure}
    \hspace*{\fill}
    \caption{Illustration of (\subref{subfig:mutation}) mutation and (\subref{subfig:crossover}) crossover used in the genetic expert policy.}
    \label{fig:gen_operators}
\end{figure}

An adequate choice of genetic operators is crucial for the expert policy. To this end, we choose the graph-based mutation and crossover proposed by \citet{jensen2019graph}, as they recently demonstrated outstanding performance for de novo molecular design. At a high-level, the genetic expert policy $\piex(x ; \mathcal{X})$ generates a molecule in two steps. First, the expert policy generates a \textit{child} molecule by applying the crossover to a pair of \textit{parent} molecules randomly drawn from $\mathcal{X}$. To be specific, two subgraphs extracted from the parents attach to form a child molecule. We use two types of operations for crossover: \texttt{non\_ring\_crossover} and \texttt{ring\_crossover}. Next, with a small probability, the expert policy mutates the child by atom-wise or bond-wise modification, e.g., adding an atom. We use seven types of operations for mutation: \texttt{atom\_deletion}, \texttt{atom\_addition}, \texttt{atom\_insertion}, \texttt{atom\_type\_change}, \texttt{bond\_order\_change}, \texttt{ring\_bond\_deletion}, and \texttt{ring\_bond\_addition}. We provide an illustration and a detailed description of the genetic operators in Figure \ref{fig:gen_operators} and Appendix \ref{sec:gen_expert}, respectively.

We also note that other choices of the improvement operators are possible for the expert policy. Especially, \citet{gao2020synthesizability} demonstrated how na\"ively applying the genetic operators may lead to proposing molecules that are unstable or unsynthesizable in real world. To consider this aspect, one may use exploration operators that considers the chemical validity of the molecules. For example, \citet{polishchuk2020crem} proposed a genetic algorithm based on ``chemically reasonable'' mutations. In addition, \citet{bradshaw2019model} proposed to generate molecules based on reaction models, which could also be used as an improvement operator in our framework. 

\textbf{Neural apprentice policy.}
We parameterize our neural apprentice policy using a long-short term memory (LSTM) network \citep{hochreiter1997long}. Moreover, the molecules are represented by a sequence of characters, i.e., the simplified molecular-input line-entry system (SMILES) \citep{weininger1988smiles} format. Under such a design, the probability $\pi(\bm{x} ; \theta)$ of a molecule $\bm{x}$ being generated from the apprentice policy is factorized into $\prod_{n=1}^{N} \pi(x_{n} | x_{1}, \ldots, x_{n-1};\theta)$. Here, $x_{1}, \ldots, x_{N}$ are the characters corresponding to canonical SMILES representation of the given molecule. 

We note that our choice of using the LSTM network to generate SMILES representations of molecules might not seem evident. Especially, molecular graph representations alternatively express the molecule, and many works have proposed new molecule-generating graph neural networks (GNNs) \citep{gomez2018automatic, jin2018junction, popova2019molecularrnn, shi2020graphaf, segler2018generating, madhawa2019graphnvp}. However, no particular GNN architecture clearly dominates over others, and recent molecular generation benchmarks \citep{brown2019guacamol, polykovskiy2018molecular} report that the LSTM network matches (or improves over) the performance of GNNs. Therefore, while searching for the best molecule-generating DNN architecture is a significant research direction, we leave it for future work. Instead, we choose the well-established LSTM architecture for the apprentice policy. 

%While architectures like graph neural networks are certainly promising, the LSTM network also seems to be competitive enough to applied in practice. 

\begin{figure}
    \centering
    \includegraphics[width=1.0\textwidth]{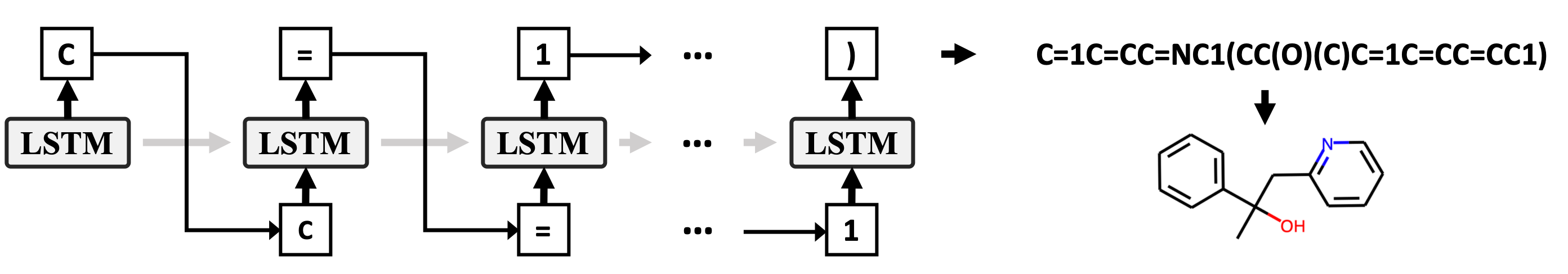}
    %\includegraphics[width=0.9\textwidth]{figure/apprentice_v0_hankook.png}
    %\includegraphics[width=0.45\textwidth]{figure/2_canonical_smiles.png}
    %\hfill
    %\includegraphics[width=0.45\textwidth]{figure/2_lstm.png}
    \caption{Illustration of the apprentice policy generating a SMILES representation of a molecule.}
    \label{fig:apprentice_policy}
    \vspace{-.15in}
\end{figure}
\textbf{Max-reward priority queues.}
Role of the max-reward priority queues $\queue$ and $\queueex$ in our framework is twofold. First, the priority queues provide highly-rewarding molecules for the expert and the apprentice policy. Furthermore, they prevent the policies from ``forgetting'' the highly-rewarding molecules observed in prior. Recent works \citep{neil2018exploring, gupta2019feedback, liang2017neural, abolafia2018neural, agarwal2019learning} have also shown similar concepts to be successful for learning in deterministic environments. 

We further elaborate on our choice of training the apprentice policy on the union of the priority queues $\queue \cup \queueex$ instead of the priority queue $\queueex$. This choice comes from how the expert policy does not always improve the apprentice policy in terms of reward, although it does in terms of exploration. Hence, it is beneficial for the apprentice policy to imitate the highly-rewarding molecules generated from both the apprentice and the expert policy. This promotes the apprentice to be trained on molecules with improved rewards.

\section{Experiments}
In this section, we report the experimental results of the proposed \algname (\ALGname) framework. To this end, we extensively compare \ALGname with the existing works, for the optimization of the penalized octanol-water partition coefficient \citep{gmezbombarelli2016automatic} and the GuacaMol benchmark \citep{brown2019guacamol}. We also provide additional experimental results in Appendix \ref{sec:additional_experiments}; these results consist of relatively straightforward tasks such as targeted optimization of the octanol-water partition coefficient and optimization of the quantitative estimate of drug-likeness (QED) \citep{bickerton2012quantifying}. For comparison, we consider a variety of existing works on de novo molecular design based on deep reinforcement learning (DRL), deep embedding optimization (DEO), genetic algorithm (GA), and deep supervised learning (DSL).\footnote{DSL-based methods train DNNs to imitate highly-rewarding molecules provided as supervisions.} We report the numbers obtained by the existing works unless stated otherwise. All of the experiments were processed using single GPU (NVIDIA RTX 2080Ti) and eight instances from a virtual CPU system (Intel Xeon E5-2630 v4). We also provide descriptions the employed baselines in Appendix \ref{sec:baselines}. A detailed illustration for the molecules generated by GEGL appears in Appendix \ref{sec:generated}.   

\textbf{Implementation details.} 
To implement GEGL, we use priority queue of fixed size $K=1024$. At each step, we sample $8192$ molecules from the apprentice and the expert policy to update the respective priority queues. Adam optimizer \citep{kingma2014adam} with learning rate of $0.001$ was used to optimize the neural network with a mini-batch of size $256$. Gradients were clipped by a norm of $1.0$. The apprentice policy is constructed using three-layered LSTM associated with hidden state of $1024$ dimensions and dropout probability of $0.2$. At each step of GEGL, the apprentice policy generates $8192$ molecules. Among the generated samples, \textit{invalid} molecules, e.g., molecules violating the valency rules, are filtered out from output of the apprentice policy. Next, the expert policy generates a molecule by selecting $8192$ pair of molecules from the priority queue to apply crossover. For each valid molecules generated from crossover operation, mutation is applied with probability of $0.01$. Similar to the apprentice policy, invalid molecules are filtered out from output of the expert policy.

\begin{table}[t]
\centering
\caption{Experimental results on the optimization of (\subref{subtab:penalized_logp}) \plogp and (\subref{subtab:penalized_logp_constrained}) \plogp with similarity constraint of $\delta$. Types of the algorithms are indicated by deep reinforcement learning (DRL), deep embedding optimization (DEO), deep supervised learning (DSL), and genetic algorithm (GA). $^{\dagger\ddagger}$We report the average and standard deviation of the objectives collected over five independent runs and 800 molecules for (\subref{subtab:penalized_logp}) and (\subref{subtab:penalized_logp_constrained}), respectively.}
%$^\ast$For GraphAF, we re-evaluate the official implementation (See Appendix \ref{sec:graphaf}).} 
\label{tab:penalized_logp}
\begin{subtable}[t]{0.48\textwidth}
\centering
\caption{\plogp}
\label{subtab:penalized_logp}
\scalebox{0.75}{
\begin{tabular}[t]{l@{\ph{0}}c@{\ph{0}}c}
\toprule
Algorithm                                           &Type       &Objective \\
\midrule
GVAE+BO {\tiny \authoryear{kusner2017grammar}}           &DEO      &\ph{0}2.87\stdv{0.06}\\
SD-VAE {\tiny \authoryear{dai2018syntaxdirected}}        &DEO      &\ph{0}3.50\stdv{0.44}\\
ORGAN {\tiny \authoryear{guimaraes2017objective}}        &DRL      &\ph{0}3.52\stdv{0.08}\\
VAE+CBO {\tiny \authoryear{griffiths2020constrained}}    &DEO      &\ph{0}4.01\ph{\stdv{0.00}}\\
ChemGE {\tiny \authoryear{yoshikawa2018population}}      &GA       &\ph{0}4.53\stdv{0.26}\\
CVAE+BO {\tiny \authoryear{gomez2018automatic}}          &DEO      &\ph{0}4.85\stdv{0.17}\\
JT-VAE {\tiny \authoryear{jin2018junction}}              &DEO      &\ph{0}4.90\stdv{0.33}\\
ChemTS {\tiny \authoryear{yang2017chemts}}               &DRL      &\ph{0}5.6\ph{0}\stdv{0.5\ph{0}}\\
GCPN {\tiny \authoryear{you2018graph}}                   &DRL      &\ph{0}7.86\stdv{0.07}\\
MRNN {\tiny \authoryear{popova2019molecularrnn}}         &DRL      &\ph{0}8.63\ph{\stdv{0.00}}\\
MolDQN {\tiny \authoryear{zhou2019optimization}}         &DRL      &11.84\ph{\stdv{0.00}}\\
GraphAF {\tiny \authoryear{shi2020graphaf}}              &DRL      &12.23\ph{\stdv{0.00}}\\
GB-GA {\tiny \authoryear{jensen2019graph}}               &GA       &15.76\stdv{5.76}\\
DA-GA {\tiny \authoryear{nigam2020augmenting}}           &GA       &20.72\stdv{3.14}\\
MSO {\tiny \authoryear{winter2019efficient}}             &DEO      &26.1\ph{0}\ph{\stdv{0.00}}\\   
PGFS {\tiny \authoryear{gottipati2020learning}}          &DRL      &27.22\ph{\stdv{0.00}}\\
GEGL$^{\dagger}$ {\tiny (Ours)}                          &DRL      &{\bf 31.40}\stdv{{\bf 0.00}}\\
\bottomrule
\end{tabular}
}
\end{subtable}
\hspace*{\fill}
\begin{subtable}[t]{0.48\textwidth}
\centering
\caption{Similarity-constrained \plogp}
\label{subtab:penalized_logp_constrained}
\vspace{.2in}
\scalebox{0.75}{
\begin{tabular}{c@{\ph{0}}l@{\ph{0}}c@{\ph{0}}c@{\ph{0}}c}
\toprule
$\delta$
&Algorithm
&Type
&Objective
&Succ. rate
\\
\midrule
\multirow{6}{*}{0.4} 
&JT-VAE {\tiny \authoryear{jin2018junction}}           &DEO     &0.84\stdv{1.45}      &0.84\\
%&GraphAF$^\ast$ {\tiny \authoryear{shi2020graphaf}}    &DRL     &1.98\stdv{1.34}      &1.00\\
&GCPN {\tiny \authoryear{you2018graph}}                &DRL     &2.49\stdv{1.30}      &1.00\\
%&MolDQN {\tiny \authoryear{zhou2019optimization}}      &DRL     &3.37\stdv{1.62}      &1.00\\
&DEFactor {\tiny \authoryear{assouel2018defactor}}     &DEO     &3.41\stdv{1.67}      &0.86\\
&VJTNN {\tiny \authoryear{jin2018learning}}            &DSL     &3.55\stdv{1.67}      &-   \\
&HierG2G {\tiny \authoryear{jin2020hierarchical}}      &DSL     &3.98\stdv{1.09}      &-   \\
%&DA-GA {\tiny \authoryear{nigam2020augmenting}}        &GA      &5.93\stdv{1.41}      &1.00\\
&GEGL$^{\ddagger}$ {\tiny(Ours)}                       &DRL     &{\bf 7.87}\stdv{{\bf 1.81}}&1.00\\
\midrule
\multirow{6}{*}{0.6} 
&JT-VAE {\tiny \authoryear{jin2018junction}}           &DEO     &0.21\stdv{0.71}      &0.47\\
&GCPN {\tiny \authoryear{you2018graph}}                &DRL     &0.79\stdv{0.63}      &1.00\\
&DEFactor {\tiny \authoryear{assouel2018defactor}}     &DEO     &1.55\stdv{1.19}      &0.73\\
%&GraphAF$^\ast$ {\tiny \authoryear{shi2020graphaf}}    &DRL     &1.68\stdv{1.22}      &0.97\\
%&MolDQN {\tiny \authoryear{zhou2019optimization}}      &DRL     &1.85\stdv{1.21}      &1.00\\
&VJTNN {\tiny \authoryear{jin2018learning}}            &DSL     &2.33\stdv{1.17}      &-   \\
&HierG2G {\tiny \authoryear{jin2020hierarchical}}      &DSL     &2.49\stdv{1.46}      &-   \\
%&DA-GA {\tiny \authoryear{nigam2020augmenting}}        &GA      &3.44\stdv{1.09}      &1.00\\
&GEGL$^{\ddagger}$ {\tiny(Ours)}                       &DRL     &{\bf 4.43}\stdv{ {\bf 1.53}}&1.00\\
\bottomrule
\end{tabular}%
}
\end{subtable}
\end{table}
\begin{figure}[t]
    \centering
    \includegraphics[width=1.0\textwidth]{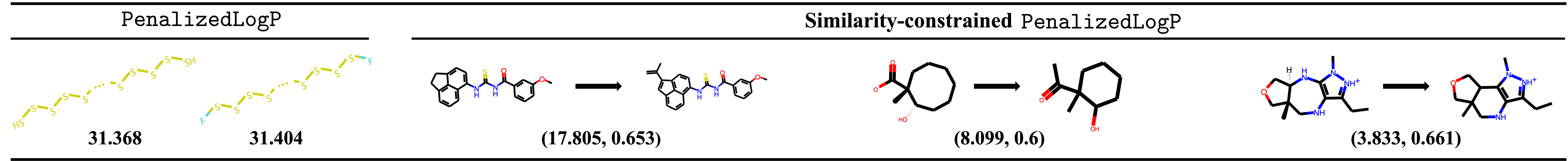}
    \caption{
    %Illustration of the highly-scoring molecules for the unconstrained and similarity-constrained \plogp optimization. 
    Illustration of the highly-scoring molecule and (reference molecule, improved molecule) for the unconstrained and similarity-constrained \plogp optimization, respectively. Below each molecule, we denote the associated objective and (objective, similarity) for the unconstrained and similarity-constrained \plogp optimization, respectively.
    }\label{fig:penalized_logp}
\end{figure}

\subsection{Optimization of penalized octanol-water partition coefficient}
\label{subsec:penalized_logp}
Comparing to the literature standard, we aim at maximizing the \textit{penalized octanol-water partition coefficient} ({\plogp}) score defined as follows:
\begin{equation*}\label{eq:plogp}
    \mathtt{PenalizedLogP}(\bm{x}) = \mathtt{LogP}(\bm{x}) - \mathtt{SyntheticAccessibility}(\bm{x}) - \mathtt{RingPenalty}(\bm{x}),
\end{equation*}
where $\mathtt{LogP}$, $\mathtt{SyntheticAccessibility}$, and $\mathtt{RingPenalty}$ correspond to the (unpenalized) octanol-water partition coefficient \citep{kwon2001handbook}, the synthetic accessibility \citep{ertl2009estimation} penalty, and the penalty for atom rings of size larger than 6. We also note that \citet{kusner2017grammar} first performed this task motivated by how the octanol-water partition coefficient plays an important role in characterizing the drug-likeness of a molecule. Following prior works \citep{jin2018junction, you2018graph}, we pretrain the apprentice policy on the ZINC dataset \citep{irwin2012zinc}. 

%Since then, it has become one of the metrics widely used to evaluate algorithms for de novo molecular design. 

\textbf{Unconstrained optimization.} 
First, we attempt to find a molecule maximizing the \plogp score as the objective without specific constraints. To this end, we run GEGL for $200$ steps while limiting the generation to molecules with at most $81$ SMILES characters (as done by \citet{jensen2019graph, nigam2020augmenting}, and \citet{yang2017chemts}). 

In Table \ref{subtab:penalized_logp}, we observe that our algorithm indeed outperforms the existing baselines by a large margin. In particular, \ALGname achieves a score of {$31.40$}, which relatively outperforms the second-best (PGFS) and the third-best (MSO) baselines by {$15\%$} and {$20\%$}, respectively. This result highlights the strong performance of \ALGname.

%This result highlights how our algorithm can outperform the existing baselines with the help of genetic exploration. 

\textbf{Constrained optimization.}
In this experiment, we follow the experimental setup proposed by \citet{jin2018junction}. Namely, for each molecule $\bm{x}_{\mathtt{ref}}$ from 800 low-\plogp-scoring molecules from the ZINC data set \citep{irwin2012zinc}, we search for a new molecule $\bm{x}$ with the maximum \plogp score while being constrained to be similar to the reference molecule $\bm{x}_{\mathtt{ref}}$. To be specific, we express this optimization as follows:
\begin{equation*}
    \max_{\bm{x}}~\mathtt{PenalizedLogP}(\bm{x})-\mathtt{PenalizedLogP}(\bm{x}_{\mathtt{ref}}), \quad \text{s.t.}\quad \mathtt{Similarity}(\bm{x}, \bm{x}_{\mathtt{ref}})\geq\delta,
\end{equation*}
where $\mathtt{Similarity}(\cdot, \cdot)$ is the Tanimoto similarity score \citep{tanimoto1958elementary}. In the experiments, we run GEGL for $50$ steps for each reference molecule $\bm{x}_{\text{ref}}$. We initialize the apprentice's priority queue $\queue$ with the reference molecule $\bm{x}_{\text{ref}}$, i.e., $\queue\gets\{\bm{x}_{\text{ref}}\}$. For this experiment, we constrain the maximum length of SMILES to be $100$. We report the above objective averaged over the $800$ molecules. We also evaluate the ``success ratio'' of the algorithms, i.e., the ratio of molecules with a positive objective. 

In Table \ref{subtab:penalized_logp_constrained}, we once again observe \ALGname to achieve superior performance to the existing algorithms. We also note that our algorithm always succeeds in improving the \plogp score of the reference molecule, i.e., the success ratio is $1.00$. 

\textbf{Generated molecules.}
We now report the molecules generated for unconstrained and constrained optimization of \plogp score in Figure \ref{fig:penalized_logp}. Notably, for unconstrained optimization, we observe that our model produces ``unrealistic'' molecules that contain a long chain of sulfurs. This symptom arise from our method exploiting the ambiguity of the \plogp score, i.e., the score spuriously assigning high values to the unrealistic molecules. Indeed, similar observations (on algorithms generating a long chain of sulfurs) have been made regardless of optimization methods for de novo molecular design. For example, \citet{shi2020graphaf}, \citet{winter2019efficient}, and \citet{nigam2020augmenting} reported molecules with similar structure when using deep reinforcement learning, deep embedding optimization, and genetic algorithm, respectively. Such an observation emphasizes why one should carefully design the procedure of de novo molecular design. %and how our method can help discover the unexpected characteristics of a given task.  

\subsection{GuacaMol benchmark}
\label{subsec:guacamol}

Next, we provide the empirical results for the GuacaMol benchmark \citep{brown2019guacamol}, designed specifically to measure the performance of de novo molecular design algorithms. It consists of 20 chemically meaningful molecular design tasks, that have been carefully designed and studied by domain-experts in the past literature \citep{brown2004graph, segler2018generating, zaliani2009second, willett1998chemical, gasteiger2006chemoinformatics}. Notably, the GuacaMol benchmark scores a set of molecules rather than a single one, to evaluate the algorithms' ability to produce diverse molecules. 
To this end, given a set of molecules $\mathcal{X}=\{\bm{x}_{s}\}_{s=1}^{|\mathcal{X}|}$ and a set of positive integers $\mathcal{S}$, tasks in the GuacaMol benchmark evaluate their score as follows:
\begin{equation*}
\mathtt{GuacaMolScore}(\mathcal{X}) \coloneqq \sum_{S\in\mathcal{S}} \sum_{s=1}^{S}\frac{r(\bm{x}_{\Pi(s)})}{S|\mathcal{S}|}, ~~\text{where} ~~r(\bm{x}_{\Pi(s)}) \geq r(\bm{x}_{\Pi(s+1)})~~\text{for}~~s = 1,\ldots,|\mathcal{X}|-1,
\end{equation*}
where $r$ is the task-specific molecule-wise score and $\Pi$ denotes a permutation that sorts the set of molecules $\mathcal{X}$ in descending order of their metrics. We further provide details on the benchmark in Appendix \ref{sec:guacamol_detail}. %Namely, the benchmark calculates one or several average score for given numbers of high-scoring molecules, and then set the final score to be the mean of such average scores.
%For instance, many of the GuacaMol benchmarks consider the average of molecule-wise scores evaluated for top-1, top-10 and top-100 molecules, i.e., $\mathcal{S}=\{1, 10, 100\}$. 

For the experiments, we initialize the apprentice policy using the weights provided by \citet{brown2019guacamol},\footnote{\url{https://github.com/BenevolentAI/guacamol_baselines}\label{fn:guacamol_baseline}} that was pretrained on the ChEMBL \citep{gaulton2012chembl} dataset. For each tasks, we run GEGL for $200$ steps. We constrain the maximum length of SMILES to be $100$.

\textbf{Benchmark results.} 
In Table \ref{tab:guacamol}, we observe that \ALGname outperforms the existing baselines by a large margin. Namely, \ALGname achieves the highest score for $19$ out of $20$ tasks. Furthermore, our algorithm perfectly solves thirteen tasks,\footnote{Every score is normalized to be in the range of $[0, 1]$.} where three of them were not known to be perfectly solved. Such a result demonstrates how our algorithm effectively produces a high-rewarding and diverse set of molecules. 

\begin{table*}[t]
\begin{center}
\caption{Experimental results for task-ids of $1, 2, \ldots,20$ from (\subref{subtab:guacamol}) the GuacaMol benchmark and (\subref{subtab:guacamol_filtered}) the GuacaMol benchmark evaluated with post-hoc filtering process. See Appendix \ref{sec:guacamol_detail} for the description of each task per id. $^{\ast}$We re-evaluate the official implementation for baselines in (\subref{subtab:guacamol_filtered}).}
\label{tab:guacamol}
\begin{subtable}[t]{0.58\textwidth}
\centering
\caption{GuacaMol}\label{subtab:guacamol}
\scalebox{0.68}{
\begin{tabular}{c@{\hspace{0.05in}}C{0.54in}@{\hspace{0.02in}}C{0.54in}@{\hspace{0.02in}}C{0.54in}@{\hspace{0.02in}}C{0.54in}@{\hspace{0.02in}}C{0.54in}@{\hspace{0.02in}}C{0.54in}@{\hspace{0.02in}}C{0.54in}@{\hspace{0.02in}}C{0.54in}@{\hspace{0.02in}}C{0.54in}@{\hspace{0.02in}}C{0.54in}}
\toprule
id
&\shortstack{ChEMBL \\ {\tiny \citep{gaulton2012chembl}}}     
&\shortstack{MCTS \\ {\tiny \citep{jensen2019graph}}}    
&\shortstack{ChemGE \\ {\tiny \citep{yoshikawa2018population}}}     
&\shortstack{HC-MLE \\ {\tiny \citep{neil2018exploring}}}	                    
&\shortstack{GB-GA \\ {\tiny \citep{jensen2019graph}}}       	    
&\shortstack{MSO \\ {\tiny \citep{winter2019efficient}}}               
&\shortstack{CReM \\ {\tiny \citep{polishchuk2020crem}}}               
&\shortstack{\ALGname \\ {\tiny (Ours)}}
\\
\midrule
\ph{0}1&{    0.505}&{    0.355}&{    0.732}&{\bf 1.000}&{\bf 1.000}&{\bf 1.000}&{\bf 1.000}&{\bf 1.000}\\
\ph{0}2&{    0.418}&{    0.311}&{    0.515}&{\bf 1.000}&{\bf 1.000}&{\bf 1.000}&{\bf 1.000}&{\bf 1.000}\\
\ph{0}3&{    0.456}&{    0.311}&{    0.598}&{\bf 1.000}&{\bf 1.000}&{\bf 1.000}&{\bf 1.000}&{\bf 1.000}\\
\ph{0}4&{    0.595}&{    0.380}&{    0.834}&{\bf 1.000}&{\bf 1.000}&{\bf 1.000}&{\bf 1.000}&{\bf 1.000}\\
\ph{0}5&{    0.719}&{    0.749}&{    0.907}&{\bf 1.000}&{\bf 1.000}&{\bf 1.000}&{\bf 1.000}&{\bf 1.000}\\
\ph{0}6&{    0.629}&{    0.402}&{    0.790}&{\bf 1.000}&{\bf 1.000}&{\bf 1.000}&{\bf 1.000}&{\bf 1.000}\\
\ph{0}7&{    0.684}&{    0.410}&{    0.829}&{    0.993}&{    0.971}&{    0.997}&{    0.966}&{\bf 1.000}\\
\ph{0}8&{    0.747}&{    0.632}&{    0.889}&{    0.879}&{    0.982}&{\bf 1.000}&{    0.940}&{\bf 1.000}\\
\ph{0}9&{    0.334}&{    0.225}&{    0.334}&{    0.438}&{    0.406}&{    0.437}&{    0.371}&{\bf 0.455}\\
10&{    0.351}&{    0.170}&{    0.380}&{    0.422}&{    0.432}&{    0.395}&{    0.434}&{\bf 0.437}\\
11&{    0.839}&{    0.784}&{    0.886}&{    0.907}&{    0.953}&{    0.966}&{    0.995}&{\bf 1.000}\\
12&{    0.817}&{    0.695}&{    0.931}&{    0.959}&{    0.998}&{\bf 1.000}&{\bf 1.000}&{\bf 1.000}\\
13&{    0.792}&{    0.616}&{    0.881}&{    0.855}&{    0.920}&{    0.931}&{\bf 0.969}&{    0.958}\\
14&{    0.575}&{    0.385}&{    0.661}&{    0.808}&{    0.792}&{    0.834}&{    0.815}&{\bf 0.882}\\
15&{    0.696}&{    0.533}&{    0.722}&{    0.894}&{    0.894}&{    0.900}&{    0.902}&{\bf 0.924}\\
16&{    0.509}&{    0.458}&{    0.689}&{    0.545}&{    0.891}&{    0.868}&{    0.763}&{\bf 0.922}\\
17&{    0.547}&{    0.488}&{    0.413}&{    0.669}&{    0.754}&{    0.764}&{    0.770}&{\bf 0.834}\\
18&{    0.259}&{    0.040}&{    0.552}&{    0.978}&{    0.990}&{    0.994}&{    0.994}&{\bf 1.000}\\
19&{    0.933}&{    0.590}&{    0.970}&{    0.996}&{\bf 1.000}&{\bf 1.000}&{\bf 1.000}&{\bf 1.000}\\
20&{    0.738}&{    0.470}&{    0.885}&{    0.998}&{\bf 1.000}&{\bf 1.000}&{\bf 1.000}&{\bf 1.000}\\
\bottomrule
\end{tabular}%
}
\end{subtable}
\hspace*{\fill}
\begin{subtable}[t]{0.39\textwidth}
\raggedright
\caption{GuacaMol with filtering}\label{subtab:guacamol_filtered}
\scalebox{0.68}{
\begin{tabular}{C{0.58in}@{\hspace{0.02in}}C{0.58in}@{\hspace{0.02in}}C{0.58in}@{\hspace{0.02in}}C{0.58in}@{\hspace{0.02in}}C{0.58in}@{\hspace{0.02in}}C{0.58in}@{\hspace{0.02in}}C{0.58in}}
\toprule
\shortstack{ChEMBL$^{\ast}$ \\ {\tiny \citep{gaulton2012chembl}}}     
&\shortstack{ChemGE$^{\ast}$ \\ {\tiny \citep{yoshikawa2018population}}}     
&\shortstack{HC-MLE$^{\ast}$ \\ {\tiny \citep{neil2018exploring}}}	                    
&\shortstack{GB-GA$^{\ast}$ \\ {\tiny \citep{jensen2019graph}}}       	    
&\shortstack{\ALGname \\ {\tiny (Ours)}}
\\
\midrule
{    0.505}&{    0.646}&{\bf 1.000}&{\bf 1.000}&{\bf 1.000}\\
{    0.260}&{    0.504}&{    0.537}&{\bf 0.837}&{    0.552}\\
{    0.456}&{    0.552}&{\bf 1.000}&{\bf 1.000}&{\bf 1.000}\\
{    0.595}&{    0.769}&{\bf 1.000}&{    0.995}&{\bf 1.000}\\
{    0.711}&{    0.959}&{\bf 1.000}&{    0.996}&{\bf 1.000}\\
{    0.632}&{    0.631}&{\bf 1.000}&{    0.996}&{\bf 1.000}\\
{    0.684}&{    0.786}&{    0.997}&{    0.960}&{\bf 1.000}\\
{    0.747}&{    0.883}&{    0.992}&{    0.823}&{\bf 1.000}\\
{    0.334}&{    0.361}&{    0.453}&{    0.402}&{\bf 0.455}\\
{    0.351}&{    0.377}&{    0.433}&{    0.420}&{\bf 0.437}\\
{    0.839}&{    0.895}&{    0.916}&{    0.914}&{\bf 1.000}\\
{    0.815}&{    0.920}&{    0.999}&{    0.905}&{\bf 1.000}\\
{    0.786}&{    0.714}&{    0.882}&{    0.530}&{\bf 0.933}\\
{    0.572}&{    0.572}&{    0.835}&{    0.780}&{\bf 0.833}\\
{    0.679}&{    0.709}&{    0.902}&{    0.889}&{\bf 0.905}\\
{    0.501}&{    0.587}&{    0.601}&{    0.634}&{\bf 0.749}\\
{    0.547}&{    0.647}&{    0.715}&{    0.698}&{\bf 0.763}\\
{    0.127}&{    0.827}&{    0.992}&{    0.789}&{\bf 1.000}\\
{    0.933}&{    0.857}&{\bf 1.000}&{\bf 0.994}&{\bf 1.000}\\
{    0.690}&{    0.964}&{\bf 1.000}&{\bf 1.000}&{\bf 1.000}\\
\bottomrule
\end{tabular}%
}
\end{subtable}
\end{center}
\end{table*}
\begin{figure}[t]
    \centering
    \includegraphics[width=1.0\textwidth]{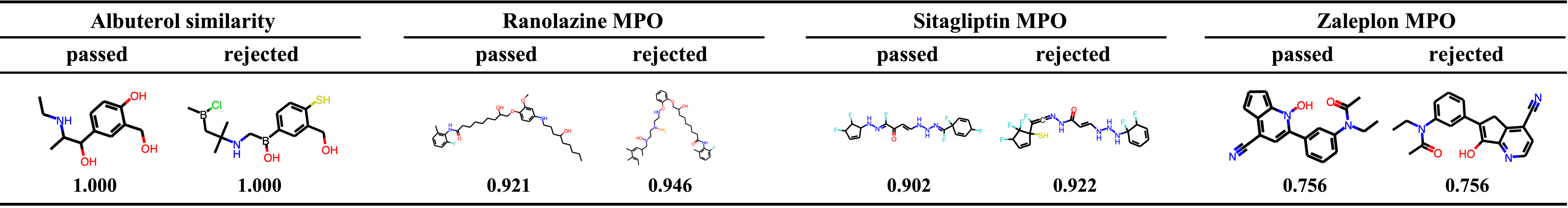}
    \caption{Illustration of the molecules generated for the GuacaMol benchmark, that have passed and rejected 
    from the filtering procedure. Below each molecule, we also denote the associated objectives.}
    \label{fig:guacamol}
\end{figure}

\textbf{Evaluation with post-hoc filtering.} 
As observed in Section \ref{subsec:penalized_logp} and prior works, de novo molecular design algorithms may lead to problematic results. For example, the generated molecules may be chemically reactive, hard to synthesize, or perceived to be ``unrealistic'' to domain experts. To consider this aspect, we evaluate our algorithm under the post-hoc filtering approach \citep{gao2020synthesizability}. Specifically, we use the expert-designed filter to reject the molecules with undesirable feature. In other words, we train the models as in Table \ref{tab:guacamol}, but only report the performance of generated molecules that pass the filter. Since the filtering procedure is post-hoc, the de novo molecular design algorithms will not be able to aggressively exploit possible ambiguities of the filtering process. 
%We further describe the filtering procedure in Appendix \ref{sec:guacamol_detail}.
%To filter out the molecules, we use the compound quality filter proposed in the GuacaMol benchmark \citet{brown2019guacamol}. 

As shown in Table \ref{subtab:guacamol_filtered}, \ALGname still outperforms the baselines even when the undesirable molecules are filtered out. This validates the ability of our algorithm to generate chemically meaningful results. Hence, we conclude that our \ALGname can be used flexibly with various choice of de novo molecular design process. 

\textbf{Generated molecules.} In Figure \ref{fig:guacamol}, we illustrate the high-scoring molecules generated by \ALGname for the Albuterol similarity, Ranolazine MPO, Sitagliptin MPO, and Zaleplon MPO tasks from the GuacaMol benchmark (corresponding to task-ids of 4, 13, 15, 16 in Table \ref{tab:guacamol}). Note that the filters used in Table \ref{subtab:guacamol_filtered} provide reasons for rejection of the molecules. For example, the high-scoring molecule from the Zaleplon MPO task was rejected due to containing a SMILES of \texttt{C=CO}, i.e., \textit{enol}. This is undesirable, as many enols have been shown to be reactive \citep{kuwajima1985reactive}. See Appendix \ref{sec:discussion_guacamol} for further explanation of the molecules being passed and rejected from the post-hoc filtering process.

\begin{figure}[t]
    \centering
    \begin{subfigure}{.245\textwidth}
    \centering
    \includegraphics[width=1.0\textwidth]{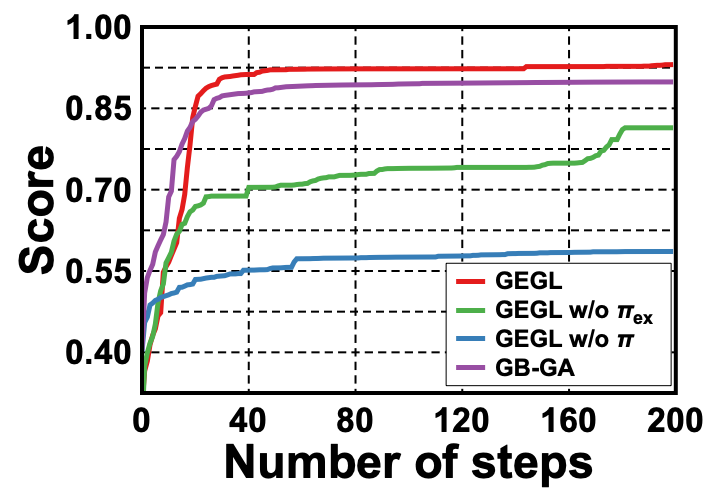}
    \caption{Sitagliptin MPO}\label{subfig:ablation0}
    \end{subfigure}
    \begin{subfigure}{.245\textwidth}
    \centering
    \includegraphics[width=1.0\textwidth]{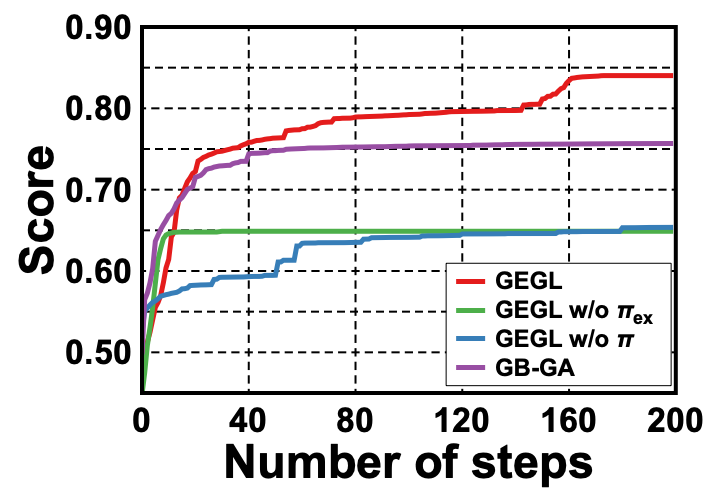}
    \caption{Zaleplon MPO}\label{subfig:ablation1}
    \end{subfigure}
    \begin{subfigure}{.245\textwidth}
    \centering
    \includegraphics[width=1.0\textwidth]{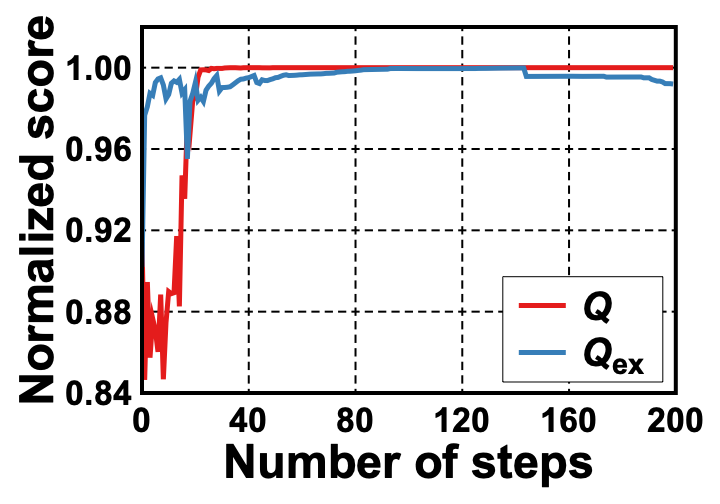}
    \caption{Sitagliptin MPO}\label{subfig:ablation2}
    \end{subfigure}
    \begin{subfigure}{.245\textwidth}
    \centering
    \includegraphics[width=1.0\textwidth]{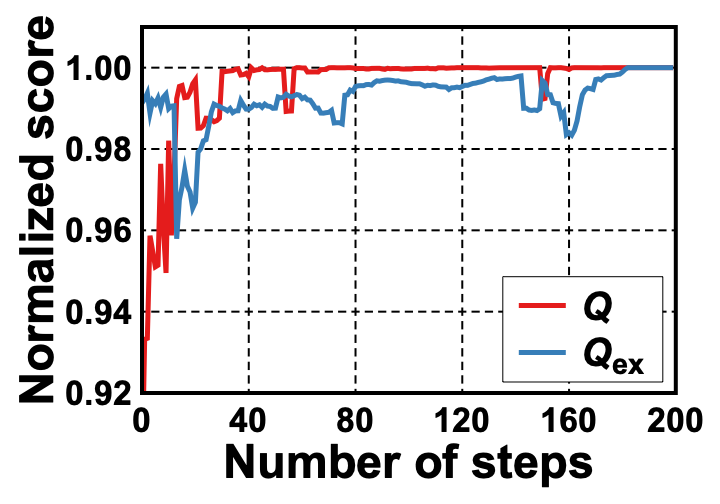}
    \caption{Zaleplon MPO}\label{subfig:ablation3}
    \end{subfigure}
    \caption{Illustration of ablation studies for (\subref{subfig:ablation0}, \subref{subfig:ablation1}) investigating contribution from DNN and genetic operator, and (\subref{subfig:ablation2}, \subref{subfig:ablation3}) separate evaluation of max-reward priority queues. }
    \label{fig:ablation}
    \vspace{-.15in}
\end{figure}
\subsection{Ablation studies}
\label{subsec:ablation}

Finally, we perform ablation studies on our algorithm to investigate the behavior of each component. To this end, we conduct experiments on the Sitagliptin MPO and Zaleplon MPO tasks from the GuacaMol benchmark. Note that Sitagliptin MPO and Zaleplon MPO tasks corresponds to task-ids of $15$ and $16$ in Table \ref{tab:guacamol}, respectively.
%To this end, we first validate that the genetic expert policy and the apprentice policy are indeed beneficial to each other for performance of the algorithm. Then we demonstrate our intriguing findings on dominance of the apprentice policy over the expert policy at later stages of executing \ALGname. 

\textbf{Contribution from the DNN and the genetic operator.} We start by inspecting the contribution of the DNN and the genetic operator in our algorithm. To this end, we compare \ALGname with (a) \ALGname without the expert policy $\pi_{\mathtt{ex}}$, (b) \ALGname without the apprentice policy $\pi$, and (c) the genetic algorithm using our improvement operators, i.e., GB-GA. To be specific, (a) trains the apprentice policy to imitate the highly-rewarding molecules generated by itself. On the other hand, (b) freezes the max-reward priority queue $\queue$ by the highly-rewarding samples from the ChEMBL dataset \citep{gaulton2012chembl}, then only updates $\queueex$ based on the expert policy. Finally, (c) is the same algorithm reported in Table \ref{tab:guacamol}, but using the hyper-parameter of our genetic expert. In Figure \ref{subfig:ablation0} and \ref{subfig:ablation1}, we observe that all the ablation algorithms perform worse than \ALGname. This result confirms that the neural apprentice policy and the genetic expert policy bring mutual benefits to our framework. 

\textbf{Separate evaluation of the max-reward priority queues.} Next, we describe behavior of the apprentice policy and the expert policy during training. To this end, we compare the GuacaMol scores for the priority queues $\queue$ and $\queueex$ that are normalized by the original \ALGname score. For example, we consider $\frac{\mathtt{GuacaMolScore}(\queue)}{\mathtt{GuacaMolScore}(\queue\cup\queueex)}$ for evaluating $\queue$ where $\mathtt{GuacaMolScore}(\cdot)$ is the GaucaMol score evaluated on a set of molecules. 

In Figure \ref{subfig:ablation2} and \ref{subfig:ablation3}, we observe that the samples collected from the genetic expert policy, i.e., $\queueex$, indeed improves over that of the apprentice policy, i.e., $\queue$ during early stages of the training. However, as the training goes on, the apprentice policy learns to generate molecules with quality higher than that of the expert policy. Since Table \ref{subfig:ablation0} and \ref{subfig:ablation1} shows that apprentice policy cannot reach the same performance without the expert policy, one may conclude that the apprentice policy effectively learns to comprise the benefits of genetic operators through learning.

\section{Conclusion}
We propose a new framework based on deep neural networks (DNNs) to solve the de novo molecular design problem. Our main idea is to enhance the training of DNN with domain knowledge, using the powerful expert-designed genetic operators to guide the training of the DNN. Through extensive experiments, our algorithm demonstrates state-of-the-art performance across a variety of tasks. We believe extending our framework to combinatorial search problems where strong genetic operators exist, e.g., biological sequence design \citep{lyngso2012frnakenstein}, program synthesis \citep{helmuth2015general}, and vehicle routing problems \citep{prins2004simple}, would be both promising and interesting. 

\newpage
\section*{Broader Impact}
\textbf{De novo molecular design.}
Our framework is likely to advance the field of de novo molecular design. In this field, successful algorithms significantly impact real-world, since the discovery of a new molecule has been the key challenge of many applications. Domain of such applications includes, but are not limited to, drug molecules \citep{griffen2018can}, organic light emitting diodes \citep{gomez2016design}, organic solar cells \citep{venkatraman2019ionic}, energetic materials \citep{jorgensen2018machine}, and electrochromic devices \citep{elton2018applying}. Improvements in these applications are beneficial to human kind in general, as they often improve the quality of human life and may broaden our knowledge of chemistry. 

%However, one should always take care in automating such molecular design processes despite its promise, especially when using a deep neural network (DNN) to synthesize molecules in a fully automated way. Currently, there is no ways of entirely explaining and regularizing the behavior of DNNs in general. Hence, DNN-based de novo molecular design might cause safety issues from unexpected behaviors and these aspect should be taken into consideration. While such an issue is resolvable by human experts inspecting the proposals of the DNN, we would also like to encourage the researchers to develop a ``safe'' framework for de novo molecular design.

\textbf{Combinatorial optimization with deep reinforcement learning.} 
In a broader sense, our framework offers a new paradigm to search over a intractably large space of combinatorial objects with DNN. In particular, our algorithm is expected to perform well for domains where genetic algorithms are powerful; this includes domains of biological sequence design \citep{lyngso2012frnakenstein}, program synthesis \citep{helmuth2015general}, and vehicle routing problems \citep{prins2004simple}. Hence, at a high-level, our work also shares the domain of applications impacted from such works. 

%More specifically, we follow the frameworks that use DNN to solve such combinatorial optimization problems. While these methods are certainly promising, we note the potential pitfall of overlooking the traditional solvers, due to the recent enthusiasm focused around deep neural networks. Furthermore, as observed in by several recent works \citep{jensen2019graph, kool2018attention}, the domain-specific methods may outperform the DNN-based methods, due to containing an important knowledge about the target problem as a strong prior. Hence, we would like to recommend the researchers to carefully apply their DNN-based methods for solving such problems, considering the domain-specific knowledge in mind.

%A deeper understanding of how the emerging deep reinforcement learning can be used for solving these important problems would bring significant gains to real-world. 
\begin{ack}
We thank Yeonghun Kang, Seonyul Kim, Jaeho Lee, Sejun Park, and Sihyun You for providing helpful feedbacks and suggestions in preparing the early version of the manuscript. This work was supported by Samsung Advanced Institute of Technology (SAIT). This work was partly supported by Institute of Information \& Communications Technology Planning \& Evaluation (IITP) grant funded by the Korea government (MSIT) (No.2019-0-00075, Artificial Intelligence Graduate School Program (KAIST)).
\end{ack}

\bibliography{reference}

\begin{thebibliography}{73}
\providecommand{\natexlab}[1]{#1}
\providecommand{\url}[1]{\texttt{#1}}
\expandafter\ifx\csname urlstyle\endcsname\relax
  \providecommand{\doi}[1]{doi: #1}\else
  \providecommand{\doi}{doi: \begingroup \urlstyle{rm}\Url}\fi

\bibitem[Hughes et~al.(2011)Hughes, Rees, Kalindjian, and
  Philpott]{hughes2011principles}
James~P Hughes, Stephen Rees, S~Barrett Kalindjian, and Karen~L Philpott.
\newblock Principles of early drug discovery.
\newblock \emph{British journal of pharmacology}, 162\penalty0 (6):\penalty0
  1239--1249, 2011.

\bibitem[Yan et~al.(2018)Yan, Barlow, Wang, Yan, Jen, Marder, and
  Zhan]{yan2018non}
Cenqi Yan, Stephen Barlow, Zhaohui Wang, He~Yan, Alex K-Y Jen, Seth~R Marder,
  and Xiaowei Zhan.
\newblock Non-fullerene acceptors for organic solar cells.
\newblock \emph{Nature Reviews Materials}, 3\penalty0 (3):\penalty0 1--19,
  2018.

\bibitem[Bohacek et~al.(1996)Bohacek, McMartin, and Guida]{bohacek1996art}
Regine~S Bohacek, Colin McMartin, and Wayne~C Guida.
\newblock The art and practice of structure-based drug design: a molecular
  modeling perspective.
\newblock \emph{Medicinal research reviews}, 16\penalty0 (1):\penalty0 3--50,
  1996.

\bibitem[Schneider and Fechner(2005)]{schneider2005computer}
Gisbert Schneider and Uli Fechner.
\newblock Computer-based de novo design of drug-like molecules.
\newblock \emph{Nature Reviews Drug Discovery}, 4\penalty0 (8):\penalty0
  649--663, 2005.

\bibitem[Schneider(2013)]{schneider2013novo}
Gisbert Schneider.
\newblock \emph{De novo molecular design}.
\newblock John Wiley \& Sons, 2013.

\bibitem[Guimaraes et~al.(2017)Guimaraes, Sanchez-Lengeling, Outeiral, Farias,
  and Aspuru-Guzik]{guimaraes2017objective}
Gabriel~Lima Guimaraes, Benjamin Sanchez-Lengeling, Carlos Outeiral, Pedro
  Luis~Cunha Farias, and Al{\'a}n Aspuru-Guzik.
\newblock Objective-reinforced generative adversarial networks (organ) for
  sequence generation models.
\newblock \emph{arXiv preprint arXiv:1705.10843}, 2017.

\bibitem[Kusner et~al.(2017)Kusner, Paige, and
  Hern{\'a}ndez-Lobato]{kusner2017grammar}
Matt~J Kusner, Brooks Paige, and Jos{\'e}~Miguel Hern{\'a}ndez-Lobato.
\newblock Grammar variational autoencoder.
\newblock In \emph{International Conference on Machine Learning}, pages
  1945--1954, 2017.

\bibitem[G{\'o}mez-Bombarelli et~al.(2018)G{\'o}mez-Bombarelli, Wei, Duvenaud,
  Hern{\'a}ndez-Lobato, S{\'a}nchez-Lengeling, Sheberla, Aguilera-Iparraguirre,
  Hirzel, Adams, and Aspuru-Guzik]{gomez2018automatic}
Rafael G{\'o}mez-Bombarelli, Jennifer~N Wei, David Duvenaud, Jos{\'e}~Miguel
  Hern{\'a}ndez-Lobato, Benjam{\'\i}n S{\'a}nchez-Lengeling, Dennis Sheberla,
  Jorge Aguilera-Iparraguirre, Timothy~D Hirzel, Ryan~P Adams, and Al{\'a}n
  Aspuru-Guzik.
\newblock Automatic chemical design using a data-driven continuous
  representation of molecules.
\newblock \emph{ACS central science}, 4\penalty0 (2):\penalty0 268--276, 2018.

\bibitem[You et~al.(2018)You, Liu, Ying, Pande, and Leskovec]{you2018graph}
Jiaxuan You, Bowen Liu, Zhitao Ying, Vijay Pande, and Jure Leskovec.
\newblock Graph convolutional policy network for goal-directed molecular graph
  generation.
\newblock In \emph{Advances in Neural Information Processing Systems}, pages
  6410--6421, 2018.

\bibitem[Liu et~al.(2018)Liu, Allamanis, Brockschmidt, and
  Gaunt]{liu2018constrained}
Qi~Liu, Miltiadis Allamanis, Marc Brockschmidt, and Alexander Gaunt.
\newblock Constrained graph variational autoencoders for molecule design.
\newblock In \emph{Advances in Neural Information Processing Systems}, pages
  7795--7804, 2018.

\bibitem[Dai et~al.(2018)Dai, Tian, Dai, Skiena, and
  Song]{dai2018syntaxdirected}
Hanjun Dai, Yingtao Tian, Bo~Dai, Steven Skiena, and Le~Song.
\newblock Syntax-directed variational autoencoder for structured data.
\newblock In \emph{International Conference on Learning Representations}, 2018.
\newblock URL \url{https://openreview.net/forum?id=SyqShMZRb}.

\bibitem[Jin et~al.(2018)Jin, Barzilay, and Jaakkola]{jin2018junction}
Wengong Jin, Regina Barzilay, and Tommi Jaakkola.
\newblock Junction tree variational autoencoder for molecular graph generation.
\newblock In \emph{International Conference on Machine Learning}, pages
  2323--2332, 2018.

\bibitem[Neil et~al.(2018)Neil, Segler, Guasch, Ahmed, Plumbley, Sellwood, and
  Brown]{neil2018exploring}
Daniel Neil, Marwin Segler, Laura Guasch, Mohamed Ahmed, Dean Plumbley, Matthew
  Sellwood, and Nathan Brown.
\newblock Exploring deep recurrent models with reinforcement learning for
  molecule design, 2018.
\newblock URL \url{https://openreview.net/forum?id=HkcTe-bR-}.

\bibitem[Assouel et~al.(2018)Assouel, Ahmed, Segler, Saffari, and
  Bengio]{assouel2018defactor}
Rim Assouel, Mohamed Ahmed, Marwin~H Segler, Amir Saffari, and Yoshua Bengio.
\newblock Defactor: Differentiable edge factorization-based probabilistic graph
  generation.
\newblock \emph{arXiv preprint arXiv:1811.09766}, 2018.

\bibitem[Bradshaw et~al.(2019)Bradshaw, Paige, Kusner, Segler, and
  Hern{\'a}ndez-Lobato]{bradshaw2019model}
John Bradshaw, Brooks Paige, Matt~J Kusner, Marwin Segler, and Jos{\'e}~Miguel
  Hern{\'a}ndez-Lobato.
\newblock A model to search for synthesizable molecules.
\newblock In \emph{Advances in Neural Information Processing Systems}, pages
  7937--7949, 2019.

\bibitem[Popova et~al.(2019)Popova, Shvets, Oliva, and
  Isayev]{popova2019molecularrnn}
Mariya Popova, Mykhailo Shvets, Junier Oliva, and Olexandr Isayev.
\newblock Molecularrnn: Generating realistic molecular graphs with optimized
  properties.
\newblock \emph{arXiv preprint arXiv:1905.13372}, 2019.

\bibitem[Winter et~al.(2019)Winter, Montanari, Steffen, Briem, No{\'e}, and
  Clevert]{winter2019efficient}
Robin Winter, Floriane Montanari, Andreas Steffen, Hans Briem, Frank No{\'e},
  and Djork-Arn{\'e} Clevert.
\newblock Efficient multi-objective molecular optimization in a continuous
  latent space.
\newblock \emph{Chemical science}, 10\penalty0 (34):\penalty0 8016--8024, 2019.

\bibitem[Zhou et~al.(2019)Zhou, Kearnes, Li, Zare, and
  Riley]{zhou2019optimization}
Zhenpeng Zhou, Steven Kearnes, Li~Li, Richard~N Zare, and Patrick Riley.
\newblock Optimization of molecules via deep reinforcement learning.
\newblock \emph{Scientific reports}, 9\penalty0 (1):\penalty0 1--10, 2019.

\bibitem[Shi et~al.(2020)Shi, Xu, Zhu, Zhang, Zhang, and Tang]{shi2020graphaf}
Chence Shi, Minkai Xu, Zhaocheng Zhu, Weinan Zhang, Ming Zhang, and Jian Tang.
\newblock Graphaf: a flow-based autoregressive model for molecular graph
  generation.
\newblock In \emph{International Conference on Learning Representations}, 2020.
\newblock URL \url{https://openreview.net/forum?id=S1esMkHYPr}.

\bibitem[Griffiths and Hern{\'a}ndez-Lobato(2020)]{griffiths2020constrained}
Ryan-Rhys Griffiths and Jos{\'e}~Miguel Hern{\'a}ndez-Lobato.
\newblock Constrained bayesian optimization for automatic chemical design using
  variational autoencoders.
\newblock \emph{Chemical Science}, 2020.

\bibitem[Jin et~al.(2020)Jin, Barzilay, and Jaakkola]{jin2020hierarchical}
Wengong Jin, Regina Barzilay, and Tommi Jaakkola.
\newblock Hierarchical generation of molecular graphs using structural motifs.
\newblock In \emph{International Conference on Machine Learning}, page to
  appear, 2020.

\bibitem[Gottipati et~al.(2020)Gottipati, Sattarov, Niu, Pathak, Wei, Liu,
  Thomas, Blackburn, Coley, Tang, et~al.]{gottipati2020learning}
Sai~Krishna Gottipati, Boris Sattarov, Sufeng Niu, Yashaswi Pathak, Haoran Wei,
  Shengchao Liu, Karam~MJ Thomas, Simon Blackburn, Connor~W Coley, Jian Tang,
  et~al.
\newblock Learning to navigate the synthetically accessible chemical space
  using reinforcement learning.
\newblock \emph{arXiv preprint arXiv:2004.12485}, 2020.

\bibitem[Yoshikawa et~al.(2018)Yoshikawa, Terayama, Sumita, Homma, Oono, and
  Tsuda]{yoshikawa2018population}
Naruki Yoshikawa, Kei Terayama, Masato Sumita, Teruki Homma, Kenta Oono, and
  Koji Tsuda.
\newblock Population-based de novo molecule generation, using grammatical
  evolution.
\newblock \emph{Chemistry Letters}, 47\penalty0 (11):\penalty0 1431--1434,
  2018.

\bibitem[Jensen(2019)]{jensen2019graph}
Jan~H Jensen.
\newblock A graph-based genetic algorithm and generative model/monte carlo tree
  search for the exploration of chemical space.
\newblock \emph{Chemical science}, 10\penalty0 (12):\penalty0 3567--3572, 2019.

\bibitem[Nigam et~al.(2020)Nigam, Friederich, Krenn, and
  Aspuru-Guzik]{nigam2020augmenting}
AkshatKumar Nigam, Pascal Friederich, Mario Krenn, and Alan Aspuru-Guzik.
\newblock Augmenting genetic algorithms with deep neural networks for exploring
  the chemical space.
\newblock In \emph{International Conference on Learning Representations}, 2020.
\newblock URL \url{https://openreview.net/forum?id=H1lmyRNFvr}.

\bibitem[Polishchuk(2020)]{polishchuk2020crem}
Pavel Polishchuk.
\newblock Crem: chemically reasonable mutations framework for structure
  generation.
\newblock \emph{Journal of Cheminformatics}, 12:\penalty0 1--18, 2020.

\bibitem[Gupta and Zou(2019)]{gupta2019feedback}
Anvita Gupta and James Zou.
\newblock Feedback gan for dna optimizes protein functions.
\newblock \emph{Nature Machine Intelligence}, 1\penalty0 (2):\penalty0
  105--111, 2019.

\bibitem[Liang et~al.(2017)Liang, Berant, Le, Forbus, and Lao]{liang2017neural}
Chen Liang, Jonathan Berant, Quoc Le, Kenneth Forbus, and Ni~Lao.
\newblock Neural symbolic machines: Learning semantic parsers on freebase with
  weak supervision.
\newblock In \emph{Proceedings of the 55th Annual Meeting of the Association
  for Computational Linguistics (Volume 1: Long Papers)}, pages 23--33, 2017.

\bibitem[Abolafia et~al.(2018)Abolafia, Norouzi, Shen, Zhao, and
  Le]{abolafia2018neural}
Daniel~A Abolafia, Mohammad Norouzi, Jonathan Shen, Rui Zhao, and Quoc~V Le.
\newblock Neural program synthesis with priority queue training.
\newblock \emph{arXiv preprint arXiv:1801.03526}, 2018.

\bibitem[Agarwal et~al.(2019)Agarwal, Liang, Schuurmans, and
  Norouzi]{agarwal2019learning}
Rishabh Agarwal, Chen Liang, Dale Schuurmans, and Mohammad Norouzi.
\newblock Learning to generalize from sparse and underspecified rewards.
\newblock In \emph{International Conference on Machine Learning}, pages
  130--140, 2019.

\bibitem[Brown et~al.(2019)Brown, Fiscato, Segler, and
  Vaucher]{brown2019guacamol}
Nathan Brown, Marco Fiscato, Marwin~HS Segler, and Alain~C Vaucher.
\newblock Guacamol: benchmarking models for de novo molecular design.
\newblock \emph{Journal of chemical information and modeling}, 59\penalty0
  (3):\penalty0 1096--1108, 2019.

\bibitem[Gao and Coley(2020)]{gao2020synthesizability}
Wenhao Gao and Connor~W Coley.
\newblock The synthesizability of molecules proposed by generative models.
\newblock \emph{Journal of Chemical Information and Modeling}, 2020.

\bibitem[Shoichet(2004)]{shoichet2004virtual}
Brian~K Shoichet.
\newblock Virtual screening of chemical libraries.
\newblock \emph{Nature}, 432\penalty0 (7019):\penalty0 862--865, 2004.

\bibitem[Bello et~al.(2016)Bello, Pham, Le, Norouzi, and
  Bengio]{bello2016neural}
Irwan Bello, Hieu Pham, Quoc~V Le, Mohammad Norouzi, and Samy Bengio.
\newblock Neural combinatorial optimization with reinforcement learning.
\newblock \emph{arXiv preprint arXiv:1611.09940}, 2016.

\bibitem[Angermueller et~al.(2020)Angermueller, Dohan, Belanger, Deshpande,
  Murphy, and Colwell]{angermueller2020model}
Christof Angermueller, David Dohan, David Belanger, Ramya Deshpande, Kevin
  Murphy, and Lucy Colwell.
\newblock Model-based reinforcement learning for biological sequence design.
\newblock In \emph{International Conference on Learning Representations}, 2020.
\newblock URL \url{https://openreview.net/forum?id=HklxbgBKvr}.

\bibitem[Brookes et~al.(2019)Brookes, Park, and
  Listgarten]{brookes2019conditioning}
David Brookes, Hahnbeom Park, and Jennifer Listgarten.
\newblock Conditioning by adaptive sampling for robust design.
\newblock In \emph{International Conference on Machine Learning}, pages
  773--782, 2019.

\bibitem[Weininger(1995)]{weininger1995method}
David Weininger.
\newblock Method and apparatus for designing molecules with desired properties
  by evolving successive populations, July~18 1995.
\newblock US Patent 5,434,796.

\bibitem[Globus et~al.(1999)Globus, Lawton, and Wipke]{globus1999automatic}
Al~Globus, John Lawton, and Todd Wipke.
\newblock Automatic molecular design using evolutionary techniques.
\newblock \emph{Nanotechnology}, 10\penalty0 (3):\penalty0 290, 1999.

\bibitem[Douguet et~al.(2000)Douguet, Thoreau, and Grassy]{douguet2000genetic}
Dominique Douguet, Etienne Thoreau, and G{\'e}rard Grassy.
\newblock A genetic algorithm for the automated generation of small organic
  molecules: Drug design using an evolutionary algorithm.
\newblock \emph{Journal of computer-aided molecular design}, 14\penalty0
  (5):\penalty0 449--466, 2000.

\bibitem[Schneider et~al.(2000)Schneider, Lee, Stahl, and
  Schneider]{schneider2000novo}
Gisbert Schneider, Man-Ling Lee, Martin Stahl, and Petra Schneider.
\newblock De novo design of molecular architectures by evolutionary assembly of
  drug-derived building blocks.
\newblock \emph{Journal of computer-aided molecular design}, 14\penalty0
  (5):\penalty0 487--494, 2000.

\bibitem[Brown et~al.(2004)Brown, McKay, Gilardoni, and
  Gasteiger]{brown2004graph}
Nathan Brown, Ben McKay, Fran{\c{c}}ois Gilardoni, and Johann Gasteiger.
\newblock A graph-based genetic algorithm and its application to the
  multiobjective evolution of median molecules.
\newblock \emph{Journal of chemical information and computer sciences},
  44\penalty0 (3):\penalty0 1079--1087, 2004.

\bibitem[Anthony et~al.(2017)Anthony, Tian, and Barber]{anthony2017thinking}
Thomas Anthony, Zheng Tian, and David Barber.
\newblock Thinking fast and slow with deep learning and tree search.
\newblock In \emph{Advances in Neural Information Processing Systems}, pages
  5360--5370, 2017.

\bibitem[Anthony et~al.(2019)Anthony, Nishihara, Moritz, Salimans, and
  Schulman]{anthony2019policy}
Thomas Anthony, Robert Nishihara, Philipp Moritz, Tim Salimans, and John
  Schulman.
\newblock Policy gradient search: Online planning and expert iteration without
  search trees.
\newblock \emph{arXiv preprint arXiv:1904.03646}, 2019.

\bibitem[Silver et~al.(2017)Silver, Schrittwieser, Simonyan, Antonoglou, Huang,
  Guez, Hubert, Baker, Lai, Bolton, et~al.]{silver2017mastering}
David Silver, Julian Schrittwieser, Karen Simonyan, Ioannis Antonoglou, Aja
  Huang, Arthur Guez, Thomas Hubert, Lucas Baker, Matthew Lai, Adrian Bolton,
  et~al.
\newblock Mastering the game of go without human knowledge.
\newblock \emph{Nature}, 550\penalty0 (7676):\penalty0 354--359, 2017.

\bibitem[Hochreiter and Schmidhuber(1997)]{hochreiter1997long}
Sepp Hochreiter and J{\"u}rgen Schmidhuber.
\newblock Long short-term memory.
\newblock \emph{Neural computation}, 9\penalty0 (8):\penalty0 1735--1780, 1997.

\bibitem[Weininger(1988)]{weininger1988smiles}
David Weininger.
\newblock Smiles, a chemical language and information system. 1. introduction
  to methodology and encoding rules.
\newblock \emph{Journal of chemical information and computer sciences},
  28\penalty0 (1):\penalty0 31--36, 1988.

\bibitem[Segler et~al.(2018)Segler, Kogej, Tyrchan, and
  Waller]{segler2018generating}
Marwin~HS Segler, Thierry Kogej, Christian Tyrchan, and Mark~P Waller.
\newblock Generating focused molecule libraries for drug discovery with
  recurrent neural networks.
\newblock \emph{ACS central science}, 4\penalty0 (1):\penalty0 120--131, 2018.

\bibitem[Madhawa et~al.(2019)Madhawa, Ishiguro, Nakago, and
  Abe]{madhawa2019graphnvp}
Kaushalya Madhawa, Katushiko Ishiguro, Kosuke Nakago, and Motoki Abe.
\newblock Graphnvp: An invertible flow model for generating molecular graphs.
\newblock \emph{arXiv preprint arXiv:1905.11600}, 2019.

\bibitem[Polykovskiy et~al.(2018)Polykovskiy, Zhebrak, Sanchez-Lengeling,
  Golovanov, Tatanov, Belyaev, Kurbanov, Artamonov, Aladinskiy, Veselov,
  et~al.]{polykovskiy2018molecular}
Daniil Polykovskiy, Alexander Zhebrak, Benjamin Sanchez-Lengeling, Sergey
  Golovanov, Oktai Tatanov, Stanislav Belyaev, Rauf Kurbanov, Aleksey
  Artamonov, Vladimir Aladinskiy, Mark Veselov, et~al.
\newblock Molecular sets (moses): a benchmarking platform for molecular
  generation models.
\newblock \emph{arXiv preprint arXiv:1811.12823}, 2018.

\bibitem[Gómez-Bombarelli et~al.(2016)Gómez-Bombarelli, Wei, Duvenaud,
  Hernández-Lobato, Sánchez-Lengeling, Sheberla, Aguilera-Iparraguirre,
  Hirzel, Adams, and Aspuru-Guzik]{gmezbombarelli2016automatic}
Rafael Gómez-Bombarelli, Jennifer~N. Wei, David Duvenaud, José~Miguel
  Hernández-Lobato, Benjamín Sánchez-Lengeling, Dennis Sheberla, Jorge
  Aguilera-Iparraguirre, Timothy~D. Hirzel, Ryan~P. Adams, and Alán
  Aspuru-Guzik.
\newblock Automatic chemical design using a data-driven continuous
  representation of molecules, 2016.

\bibitem[Bickerton et~al.(2012)Bickerton, Paolini, Besnard, Muresan, and
  Hopkins]{bickerton2012quantifying}
G~Richard Bickerton, Gaia~V Paolini, J{\'e}r{\'e}my Besnard, Sorel Muresan, and
  Andrew~L Hopkins.
\newblock Quantifying the chemical beauty of drugs.
\newblock \emph{Nature chemistry}, 4\penalty0 (2):\penalty0 90, 2012.

\bibitem[Kingma and Ba(2014)]{kingma2014adam}
Diederik~P Kingma and Jimmy Ba.
\newblock Adam: A method for stochastic optimization.
\newblock \emph{arXiv preprint arXiv:1412.6980}, 2014.

\bibitem[Yang et~al.(2017)Yang, Zhang, Yoshizoe, Terayama, and
  Tsuda]{yang2017chemts}
Xiufeng Yang, Jinzhe Zhang, Kazuki Yoshizoe, Kei Terayama, and Koji Tsuda.
\newblock Chemts: an efficient python library for de novo molecular generation.
\newblock \emph{Science and technology of advanced materials}, 18\penalty0
  (1):\penalty0 972--976, 2017.

\bibitem[Jin et~al.(2019)Jin, Yang, Barzilay, and Jaakkola]{jin2018learning}
Wengong Jin, Kevin Yang, Regina Barzilay, and Tommi Jaakkola.
\newblock Learning multimodal graph-to-graph translation for molecule
  optimization.
\newblock In \emph{International Conference on Learning Representations}, 2019.
\newblock URL \url{https://openreview.net/forum?id=B1xJAsA5F7}.

\bibitem[Kwon(2001)]{kwon2001handbook}
Younggil Kwon.
\newblock \emph{Handbook of essential pharmacokinetics, pharmacodynamics and
  drug metabolism for industrial scientists}.
\newblock Springer Science \& Business Media, 2001.

\bibitem[Ertl and Schuffenhauer(2009)]{ertl2009estimation}
Peter Ertl and Ansgar Schuffenhauer.
\newblock Estimation of synthetic accessibility score of drug-like molecules
  based on molecular complexity and fragment contributions.
\newblock \emph{Journal of cheminformatics}, 1\penalty0 (1):\penalty0 8, 2009.

\bibitem[Irwin et~al.(2012)Irwin, Sterling, Mysinger, Bolstad, and
  Coleman]{irwin2012zinc}
John~J Irwin, Teague Sterling, Michael~M Mysinger, Erin~S Bolstad, and Ryan~G
  Coleman.
\newblock Zinc: a free tool to discover chemistry for biology.
\newblock \emph{Journal of chemical information and modeling}, 52\penalty0
  (7):\penalty0 1757--1768, 2012.

\bibitem[Tanimoto(1958)]{tanimoto1958elementary}
Taffee~T Tanimoto.
\newblock Elementary mathematical theory of classification and prediction.
\newblock 1958.

\bibitem[Zaliani et~al.(2009)Zaliani, Boda, Seidel, Herwig, Schwab, Gasteiger,
  Clau{\ss}en, Lemmen, Degen, P{\"a}rn, et~al.]{zaliani2009second}
Andrea Zaliani, Krisztina Boda, Thomas Seidel, Achim Herwig, Christof~H Schwab,
  Johann Gasteiger, Holger Clau{\ss}en, Christian Lemmen, J{\"o}rg Degen, Juri
  P{\"a}rn, et~al.
\newblock Second-generation de novo design: a view from a medicinal chemist
  perspective.
\newblock \emph{Journal of computer-aided molecular design}, 23\penalty0
  (8):\penalty0 593--602, 2009.

\bibitem[Willett et~al.(1998)Willett, Barnard, and Downs]{willett1998chemical}
Peter Willett, John~M Barnard, and Geoffrey~M Downs.
\newblock Chemical similarity searching.
\newblock \emph{Journal of chemical information and computer sciences},
  38\penalty0 (6):\penalty0 983--996, 1998.

\bibitem[Gasteiger and Engel(2006)]{gasteiger2006chemoinformatics}
Johann Gasteiger and Thomas Engel.
\newblock \emph{Chemoinformatics: a textbook}.
\newblock John Wiley \& Sons, 2006.

\bibitem[Gaulton et~al.(2012)Gaulton, Bellis, Bento, Chambers, Davies, Hersey,
  Light, McGlinchey, Michalovich, Al-Lazikani, et~al.]{gaulton2012chembl}
Anna Gaulton, Louisa~J Bellis, A~Patricia Bento, Jon Chambers, Mark Davies,
  Anne Hersey, Yvonne Light, Shaun McGlinchey, David Michalovich, Bissan
  Al-Lazikani, et~al.
\newblock Chembl: a large-scale bioactivity database for drug discovery.
\newblock \emph{Nucleic acids research}, 40\penalty0 (D1):\penalty0
  D1100--D1107, 2012.

\bibitem[Kuwajima and Nakamura(1985)]{kuwajima1985reactive}
Isao Kuwajima and Eiichi Nakamura.
\newblock Reactive enolates from enol silyl ethers.
\newblock \emph{Accounts of Chemical Research}, 18\penalty0 (6):\penalty0
  181--187, 1985.

\bibitem[Lyngs{\o} et~al.(2012)Lyngs{\o}, Anderson, Sizikova, Badugu, Hyland,
  and Hein]{lyngso2012frnakenstein}
Rune~B Lyngs{\o}, James~WJ Anderson, Elena Sizikova, Amarendra Badugu, Tomas
  Hyland, and Jotun Hein.
\newblock Frnakenstein: multiple target inverse rna folding.
\newblock \emph{BMC bioinformatics}, 13\penalty0 (1):\penalty0 260, 2012.

\bibitem[Helmuth and Spector(2015)]{helmuth2015general}
Thomas Helmuth and Lee Spector.
\newblock General program synthesis benchmark suite.
\newblock In \emph{Proceedings of the 2015 Annual Conference on Genetic and
  Evolutionary Computation}, pages 1039--1046, 2015.

\bibitem[Prins(2004)]{prins2004simple}
Christian Prins.
\newblock A simple and effective evolutionary algorithm for the vehicle routing
  problem.
\newblock \emph{Computers \& Operations Research}, 31\penalty0 (12):\penalty0
  1985--2002, 2004.

\bibitem[Griffen et~al.(2018)Griffen, Dossetter, Leach, and
  Montague]{griffen2018can}
Edward~J Griffen, Alexander~G Dossetter, Andrew~G Leach, and Shane Montague.
\newblock Can we accelerate medicinal chemistry by augmenting the chemist with
  big data and artificial intelligence?
\newblock \emph{Drug discovery today}, 2018.

\bibitem[G{\'o}mez-Bombarelli et~al.(2016)G{\'o}mez-Bombarelli,
  Aguilera-Iparraguirre, Hirzel, Duvenaud, Maclaurin, Blood-Forsythe, Chae,
  Einzinger, Ha, Wu, et~al.]{gomez2016design}
Rafael G{\'o}mez-Bombarelli, Jorge Aguilera-Iparraguirre, Timothy~D Hirzel,
  David Duvenaud, Dougal Maclaurin, Martin~A Blood-Forsythe, Hyun~Sik Chae,
  Markus Einzinger, Dong-Gwang Ha, Tony Wu, et~al.
\newblock Design of efficient molecular organic light-emitting diodes by a
  high-throughput virtual screening and experimental approach.
\newblock \emph{Nature materials}, 15\penalty0 (10):\penalty0 1120--1127, 2016.

\bibitem[Venkatraman et~al.(2019)Venkatraman, Evjen, and
  Chellappan~Lethesh]{venkatraman2019ionic}
Vishwesh Venkatraman, Sigvart Evjen, and Kallidanthiyil Chellappan~Lethesh.
\newblock The ionic liquid property explorer: An extensive library of
  task-specific solvents.
\newblock \emph{Data}, 4\penalty0 (2):\penalty0 88, 2019.

\bibitem[J{\o}rgensen et~al.(2018)J{\o}rgensen, Mesta, Shil,
  Garc{\'\i}a~Lastra, Jacobsen, Thygesen, and Schmidt]{jorgensen2018machine}
Peter~Bj{\o}rn J{\o}rgensen, Murat Mesta, Suranjan Shil, Juan~Maria
  Garc{\'\i}a~Lastra, Karsten~Wedel Jacobsen, Kristian~Sommer Thygesen, and
  Mikkel~N Schmidt.
\newblock Machine learning-based screening of complex molecules for polymer
  solar cells.
\newblock \emph{The Journal of chemical physics}, 148\penalty0 (24):\penalty0
  241735, 2018.

\bibitem[Elton et~al.(2018)Elton, Boukouvalas, Butrico, Fuge, and
  Chung]{elton2018applying}
Daniel~C Elton, Zois Boukouvalas, Mark~S Butrico, Mark~D Fuge, and Peter~W
  Chung.
\newblock Applying machine learning techniques to predict the properties of
  energetic materials.
\newblock \emph{Scientific reports}, 8\penalty0 (1):\penalty0 1--12, 2018.

\bibitem[Gabrielson(2018)]{gabrielson2018scifinder}
Stephen~Walter Gabrielson.
\newblock Scifinder.
\newblock \emph{Journal of the Medical Library Association: JMLA}, 106\penalty0
  (4):\penalty0 588, 2018.

\bibitem[Wishart et~al.(2018)Wishart, Feunang, Guo, Lo, Marcu, Grant, Sajed,
  Johnson, Li, Sayeeda, et~al.]{wishart2018drugbank}
David~S Wishart, Yannick~D Feunang, An~C Guo, Elvis~J Lo, Ana Marcu, Jason~R
  Grant, Tanvir Sajed, Daniel Johnson, Carin Li, Zinat Sayeeda, et~al.
\newblock Drugbank 5.0: a major update to the drugbank database for 2018.
\newblock \emph{Nucleic acids research}, 46\penalty0 (D1):\penalty0
  D1074--D1082, 2018.

\end{thebibliography}
\bibliographystyle{unsrtnat}

\newpage

\appendix

\section{Details of the genetic operators}
\label{sec:gen_expert}

In this paper, we use crossover and mutation proposed by \citet{jensen2019graph} for exploring the chemical space. 

\textbf{Crossover.} The crossover randomly applies either \texttt{non\_ring\_crossover} or \texttt{ring\_crossover}, with equal probability. This generates two (possibly invalid) child molecules. If both the child molecules are invalid, e.g., violating the valency rules, the crossover is re-applied with limited number of trials. If valid molecules exist, the we choose one of them randomly. In the following, we provide further details on the \texttt{non\_ring\_crossover} and \texttt{ring\_crossover}. 

\begin{itemize}[leftmargin=0.6in]
\item[\textbf{a.}] The \texttt{non\_ring\_crossover} cuts an arbitrary edge, which does not belong to ring, of two parent molecules, and then attach the subgraphs from different parent molecules. 

\item[\textbf{b.}] The \texttt{ring\_crossover} cuts two edges in an arbitrary ring, and attach the subgraphs from different parent molecules. 
%If the cut failed to create two disconnected subgraphs from a \textit{parent} molecule, discard it and re-try the \textit{ring crossover} with respect to the original \textit{parent} molecules.
\end{itemize}

\textbf{Mutation.} The mutation randomly applies one of the seven different ways for modifying a molecule: \texttt{atom\_deletion}, \texttt{atom\_addition}, \texttt{atom\_insertion}, \texttt{atom\_type\_change}, \texttt{ring\_bond\_deletion}, \texttt{ring\_bond\_addition}, and \texttt{bond\_order\_change}. %One of the seven ways is chosen with predefined probability at each mutation, i.e., (0.15, 0.14, 0.14, 0.14, 0.14, 0.14, 0.15). 
% A target atom or a target bond, which is perturbed by the mutation in a given molecule, is randomly selected. I
After mutation, if the modified molecule is not valid, we discard it and re-apply mutation. Details of seven different ways of modifying a molecule are as follows. 
 
 \begin{itemize}[leftmargin=0.6in]
 \item[\textbf{a.}] The \texttt{atom\_deletion} removes a single atom and rearrange neighbor molecules with minimal deformation from the original molecule.
 
 \item[\textbf{b.}] The \texttt{atom\_addition} connects a new atom to a single atom. 
 
 \item[\textbf{c.}] The \texttt{atom\_insertion} puts an atom between two atoms. 
  
 \item[\textbf{d.}] The \texttt{atom\_type\_change} newly changes a type of an atom.

 \item[\textbf{e.}] The \texttt{bond\_order\_change} alters the type of a bond. 
 
 \item[\textbf{f.}] The \texttt{ring\_bond\_deletion} cuts a bond from ring. 
 
 \item[\textbf{g.}] The \texttt{ring\_bond\_addition} creates a ``shortcut'' between two connected atoms. 
\end{itemize}

\newpage

\section{Additional experiments}
\label{sec:additional_experiments}
\begin{table}[h!]
\centering
\caption{Experiment results for quantitative estimate of drug-likeness (QED) task.} 
\label{tab:qed}
\begin{tabular}[t]{l@{\ph{0}}c@{\ph{0}}c}
\toprule
Algorithm                                           &Type       &Objective \\
\midrule
MCTS {\tiny \authoryear{jensen2019graph}}                &MCTS     &0.851\\
ORGAN {\tiny \authoryear{guimaraes2017objective}}        &DRL      &0.896\\
JT-VAE {\tiny \authoryear{jin2018junction}}              &DEO      &0.925\\
ChemGE {\tiny \authoryear{yoshikawa2018population}}      &GA       &\textbf{0.948}\\
GCPN {\tiny \authoryear{you2018graph}}                   &DRL      &\textbf{0.948}\\
MRNN {\tiny \authoryear{popova2019molecularrnn}}         &DRL      &\textbf{0.948}\\
MolDQN {\tiny \authoryear{zhou2019optimization}}         &DRL      &\textbf{0.948}\\
GraphAF {\tiny \authoryear{shi2020graphaf}}              &DRL      &\textbf{0.948}\\
GB-GA {\tiny \authoryear{jensen2019graph}}               &GA       &\textbf{0.948}\\
HC-MLE {\tiny \authoryear{jensen2019graph}}              &DRL      &\textbf{0.948}\\
MSO {\tiny \authoryear{winter2019efficient}}             &DEO      &\textbf{0.948}\\
GEGL$^{\dagger}$ {\tiny (Ours)}                          &DRL      &\textbf{0.948}\\
\bottomrule
\end{tabular}
\end{table}
\begin{table*}[h!]
\begin{center}
\centering
\caption{Experimental results on relatively straight-forward tasks from the Guacamol benchmark.}
\label{tab:guacamol_trivial}
\begin{tabular}{l@{\hspace{0.05in}}C{0.54in}@{\hspace{0.02in}}C{0.54in}@{\hspace{0.02in}}C{0.54in}@{\hspace{0.02in}}C{0.54in}@{\hspace{0.02in}}C{0.54in}@{\hspace{0.02in}}C{0.54in}@{\hspace{0.02in}}C{0.54in}@{\hspace{0.02in}}C{0.54in}@{\hspace{0.02in}}C{0.54in}@{\hspace{0.02in}}C{0.54in}}
\toprule
Task
&\shortstack{ChEMBL \\ {\tiny \citep{gaulton2012chembl}}}     
&\shortstack{MCTS \\ {\tiny \citep{jensen2019graph}}}    
&\shortstack{ChemGE \\ {\tiny \citep{yoshikawa2018population}}}     
&\shortstack{HC-MLE \\ {\tiny \citep{neil2018exploring}}}	                    
&\shortstack{GB-GA \\ {\tiny \citep{jensen2019graph}}}
&\shortstack{\ALGname \\ {\tiny (Ours)}}               
\\
\midrule
logP (target: -1.0)             &{\bf 1.000}&{\bf 1.000}&{\bf 1.000}&{\bf 1.000}&{\bf 1.000}&{\bf 1.000}\\
logP (target: 8.0)              &{\bf 1.000}&{\bf 1.000}&{\bf 1.000}&{\bf 1.000}&{\bf 1.000}&{\bf 1.000}\\
TPSA (target: 150.0)            &{\bf 1.000}&{\bf 1.000}&{\bf 1.000}&{\bf 1.000}&{\bf 1.000}&{\bf 1.000}\\
CNS MPO                         &{\bf 1.000}&{\bf 1.000}&{\bf 1.000}&{\bf 1.000}&{\bf 1.000}&{\bf 1.000}\\
C$_{7}$H$_{8}$N$_{2}$O$_{2}$    &{    0.972}&{    0.851}&{    0.992}&{\bf 1.000}&{    0.993}&{\bf 1.000}\\
Pioglitazone MPO                &{    0.982}&{    0.941}&{\bf 1.000}&{    0.993}&{    0.998}&{    0.999}\\
\bottomrule
\end{tabular}
\end{center}
\end{table*}

In this section, we provide additional experimental results for the tasks in the GuacaMol benchmark \citet{brown2019guacamol} that were determined to be relatively more straight-forward than other tasks. To this end, we first report our result for unconstrained optimization of the quantitative estimate of drug-likeness (QED) \citep{bickerton2012quantifying} in Table \ref{tab:qed}. Next, we evaluate GEGL on other tasks from the GaucaMol benchmark in Table \ref{tab:guacamol_trivial}. In the Table \ref{tab:qed} and \ref{tab:guacamol_trivial}, we observe GEGL to achieve the highest scores for six out of seven tasks. See Appendix \ref{sec:guacamol_detail} for details on the tasks considered in this section.

\newpage

%For the experiments regarding the \plogp, we pre-train the apprentice policy on the ZINC dataset \citep{irwin2012zinc}. For the experiments on the GuacaMol benchmark, we pre-train the apprentice policy on the ChemBLE dataset modified by \citet{brown2019guacamol}. 

%For unconstrained optimization of \plogp and the GaucaMol benchmark, we train the apprentice policy for $1000$ steps. 

%For the similarity-constrained \plogp optimization, we train the apprentice policy for $50$ steps for each of $800$ molecules. Furthermore, we initialize the apprentice's max-reward priority queue $\queue$ by the reference molecule.

\newpage
\section{Baselines for de novo molecular design}
\label{sec:baselines}
In this section, we briefly describe the algorithms we used as baselines for evaluating our algorithm. 
\begin{itemize}[leftmargin=0.6in]
\item[\textbf{1.}] GVAE+BO \citep{kusner2017grammar} trains a \textit{grammar} variational autoencoder and applies Bayesian optimization on its embedding space.
\item[\textbf{2.}] SD-VAE \citep{dai2018syntaxdirected} propose a \textit{syntax directed} variational autoencoder and applies Bayesian optimization on its embedding space.
\item[\textbf{3.}] ORGAN \citep{guimaraes2017objective} trains a generative adversarial network along with reinforcement learning for maximizing the object.
\item[\textbf{4.}] VAE+CBO \citep{griffiths2020constrained} trains a molecule-generating variational autoencoder and applies constrained Bayesian optimization on its embedding space.
\item[\textbf{5.}] CVAE+BO \citep{gomez2018automatic} trains a \textit{constrained} variational autoencoder and applies Bayesian optimization on its embedding space.
\item[\textbf{6.}] JT-VAE \citep{jin2018junction} trains a \textit{junction tree} variational autoencoder and applies Bayesian optimization on its embedding space.
\item[\textbf{7.}] ChemTS \citep{yang2017chemts} proposes a SMILES-based genetic algorithm.
\item[\textbf{8.}] GCPN \citep{you2018graph} trains a policy parameterized with graph convolutional network with reinforcement learning to generate highly-rewarding molecules. The reward is defined as a linear combination of the desired property of the molecule and a discriminator term for generating ``realistic'' molecules. 
\item[\textbf{9.}] Molecular recurrent neural network (MRNN) \citep{popova2019molecularrnn} trains a recurrent neural network using reinforcement learning. 
\item[\textbf{10.}] MolDQN \citep{zhou2019optimization} trains a molecular fingerprint-based policy with reinforcement learning.
\item[\textbf{11.}] GraphAF \citep{shi2020graphaf} trains a auto-regressive flow model with reinforcement learning.
\item[\textbf{12.}] GB-GA \citep{jensen2019graph} proposes a graph-based genetic algorithm.
\item[\textbf{12.}] MCTS \citep{jensen2019graph} proposes a Monte Carlo tree search algorithm.
\item[\textbf{13.}] MSO \citep{winter2019efficient} applies particle swarm optimization on the embedding space of a variational autoencoder. 
\item[\textbf{14.}] DEFactor \citep{assouel2018defactor} trains a variational autoencoder where computational efficiency was enhanced with differentiable edge variables.
\item[\textbf{15.}] VJTNN \citep{jin2018learning} trains a graph-to-graph model with supervised learning on the highly-rewarding set of molecules. 
\item[\textbf{16.}] HC-MLE \citep{neil2018exploring} proposes a ``hill-climbing'' variant of reinforcement learning to train a recurrent network.
\item[\textbf{17.}] CReM \citep{polishchuk2020crem} proposes a genetic algorithm based on ``chemically reasonable'' genetic operator to generate a set of chemically reasonable molecules.  
\item[\textbf{18.}] HierG2G \citep{jin2020hierarchical} trains a graph-to-graph, hierarchical generative model with supervised learning on the highly-rewarding set of molecules.
\item[\textbf{19.}] DA-GA \citep{nigam2020augmenting} proposes a genetic algorithm based on its fitness function augmented with a discriminator which assigns higher scores to ``novel'' elements.
\item[\textbf{20.}] PGFS \citep{gottipati2020learning} proposes a reinforcement learning algorithm that generates molecules from deciding the synthesis route given the set of available reactants.
\end{itemize}

\newpage
\section{Additional Illustration of the Generated Molecules}
\label{sec:generated}

\begin{figure}[h!]
    \centering
    \includegraphics[width=\textwidth]{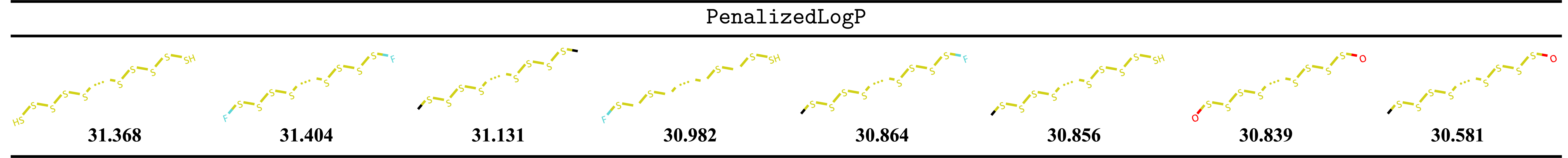}
    \caption{Additional illustration of highly-scoring molecules for the unconstrained \plogp optimization. Below each molecule, we denote the associated objective.}
    \label{fig:unconstrained_plogp}
\end{figure}

\begin{figure}[h!]
    \centering
    \includegraphics[width=\textwidth]{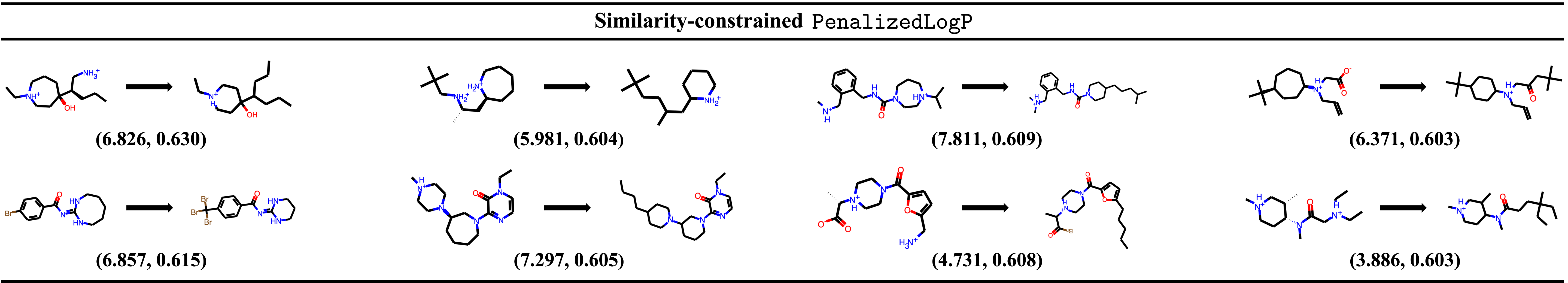}
    \caption{Additional illustration of highly-scoring molecules for the similarity-constrained \plogp optimization. Below each molecule, we denote the associated (objective, similarity).}
    \label{fig:constrained_plogp}
\end{figure}

\begin{figure}[h!]
    \centering
    \includegraphics[width=\textwidth]{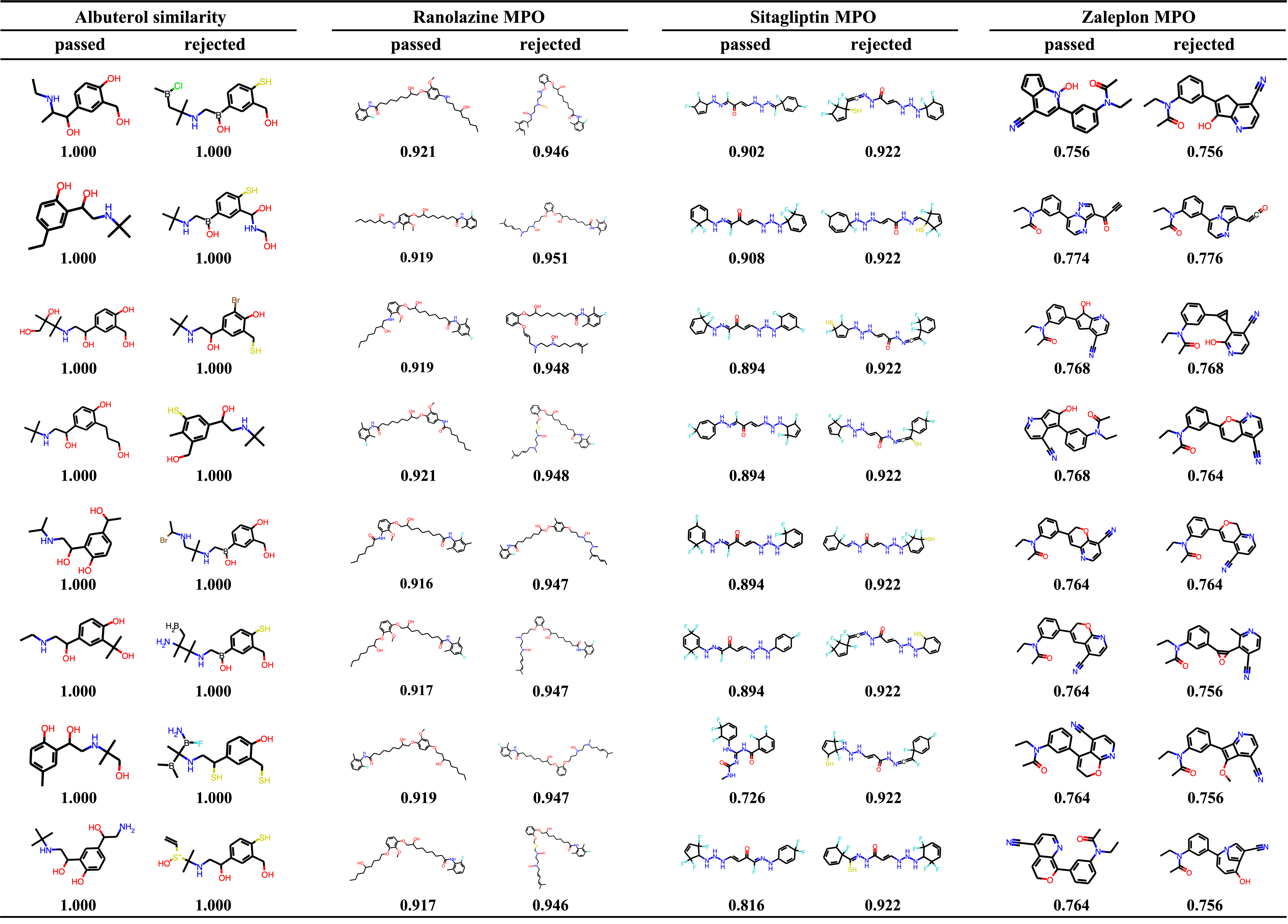}
    \caption{Additional illustration of the molecules generated for the GuacaMol benchmark, that have passed and rejected from the filtering procedure. Below each molecule, we denote the associated objective.}
    \label{fig:app_guacamol}
\end{figure}

\newpage
\section{Details on the GuacaMol benchmark}
\label{sec:guacamol_detail}
In this section, we further provide details on the tasks considered in the GuacaMol benchmark. 
\begin{itemize}
\item[$\bullet$] \textbf{Rediscovery.} The \{celecoxib, troglitazone, thiothixene\} rediscovery tasks desire the molecules to be as similar as possible to the target molecules. The algorithms achieve the maximum score when they produce a molecule identical to the target molecule. These benchmarks have been studied by \citet{zaliani2009second} and \citet{segler2018generating}. 

\item[$\bullet$] \textbf{Similarity.} The \{aripiprazole, albuterol, mestranol\} similarity tasks also aim at finding a molecule similar to the target molecule. It is different from the rediscovery task since the metric is evaluated over multiple molecules. This similarity metric has been studied by \citet{willett1998chemical}.

\item[$\bullet$] \textbf{Isometry.} The tasks of C$_{11}$H$_{24}$,  C$_{9}$H$_{10}$N$_{2}$O$_{2}$PF$_{2}$Cl, C$_{7}$H$_{8}$N$_{2}$O$_{2}$ attempt to find molecules with the target formula. The algorithms achieve an optimal score when they find all of the possible isometric molecules for the tasks.

\item[$\bullet$] \textbf{Median molecules.} The median molecule tasks searches for molecules that are simultaneously similar to a pair of molecules. This previously has been studied by \citet{brown2004graph}. 

\item[$\bullet$] \textbf{Multi property optimization.} The \{osimertinib, fexofenadine, ranolazine, perindopril, amlodipine, CNS, Pioglitazone\} multi property optimization (MPO) tasks attempt to fine-tune the structural and physicochemical properties of known drug molecules. For example, 
for the sitagliptin MPO benchmark, the models must generate molecules that are as dissimilar to sitagliptin as possible, while keeping some of its properties.

\item[$\bullet$] \textbf{Other tasks.} The valsartan SMARTS benchmark targets molecules containing a SMARTS pattern related to valsartan while being characterized by physicochemical properties corresponding to the sitagliptin molecule. Next, the scaffold Hop and decorator Hop benchmarks aim to maximize the similarity to a SMILES strings, while keeping or excluding specific SMARTS patterns, mimicking the tasks of changing the scaffold of a compound while keeping specific substituents, and keeping a scaffold fixed while changing the substitution pattern. The LogP tasks aim at generating molecules with the targeted value of octanol-water partition coefficient. The TPSA task attempts to find a molecule with the targeted value of topological polar surface area. The QED task aims at maximizing the quantitative estimate of drug-likeness score.
\end{itemize}

%\section{Description of GEGL without expert or apprentice policy}
%\label{sec:GEGL_ablated}

\newpage
\section{Discussion on GuacaMol molecules}
\label{sec:discussion_guacamol}
\begin{figure}[H]
    \centering
    \includegraphics[width=0.8\textwidth]{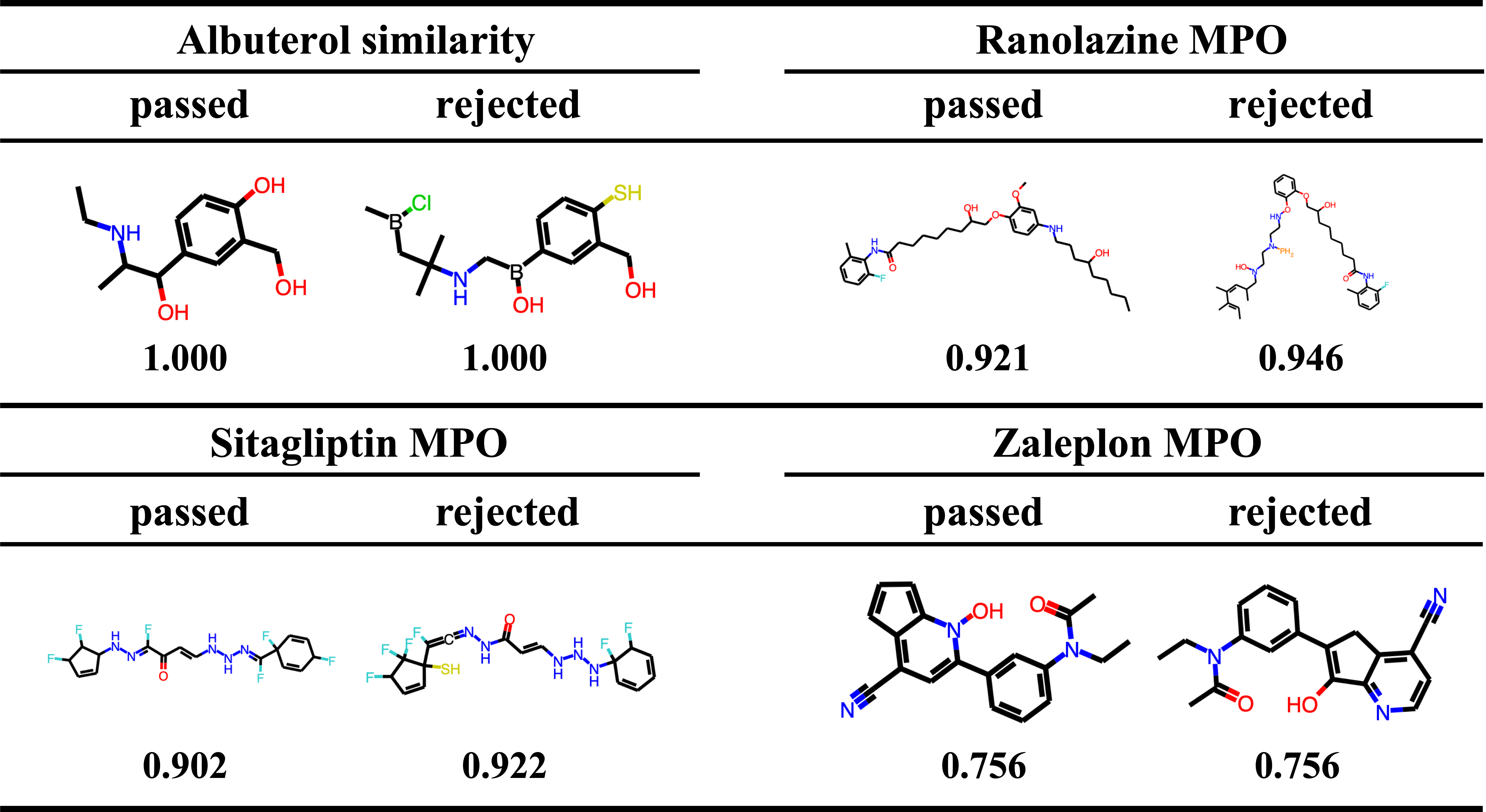}
    \caption{Illustration of the molecules generated for the GuacaMol benchmark, that have passed and rejected 
    from the filtering procedure. Below each molecule, we also denote the associated objectives.}
    \label{fig:guacamol_duplicate}
\end{figure}
\begin{figure}[H]
    \centering
    \begin{subfigure}{.3\textwidth}
    \centering
    \includegraphics[width=1.0\textwidth]{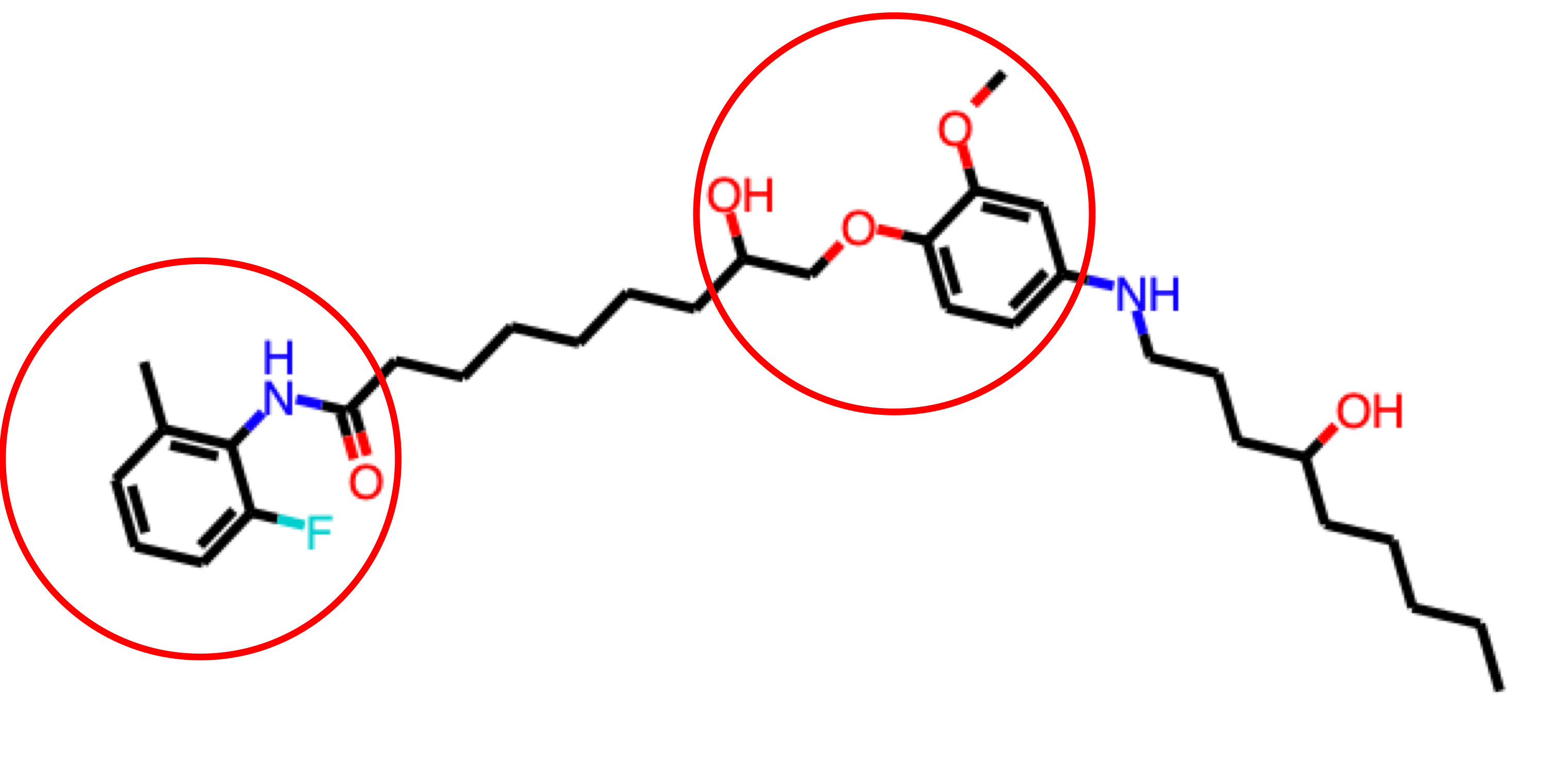}
    \caption{}
    \label{subfig:ranolazine_passed}
    \end{subfigure}
    \begin{subfigure}{.3\textwidth}
    \centering
    \includegraphics[width=1.0\textwidth]{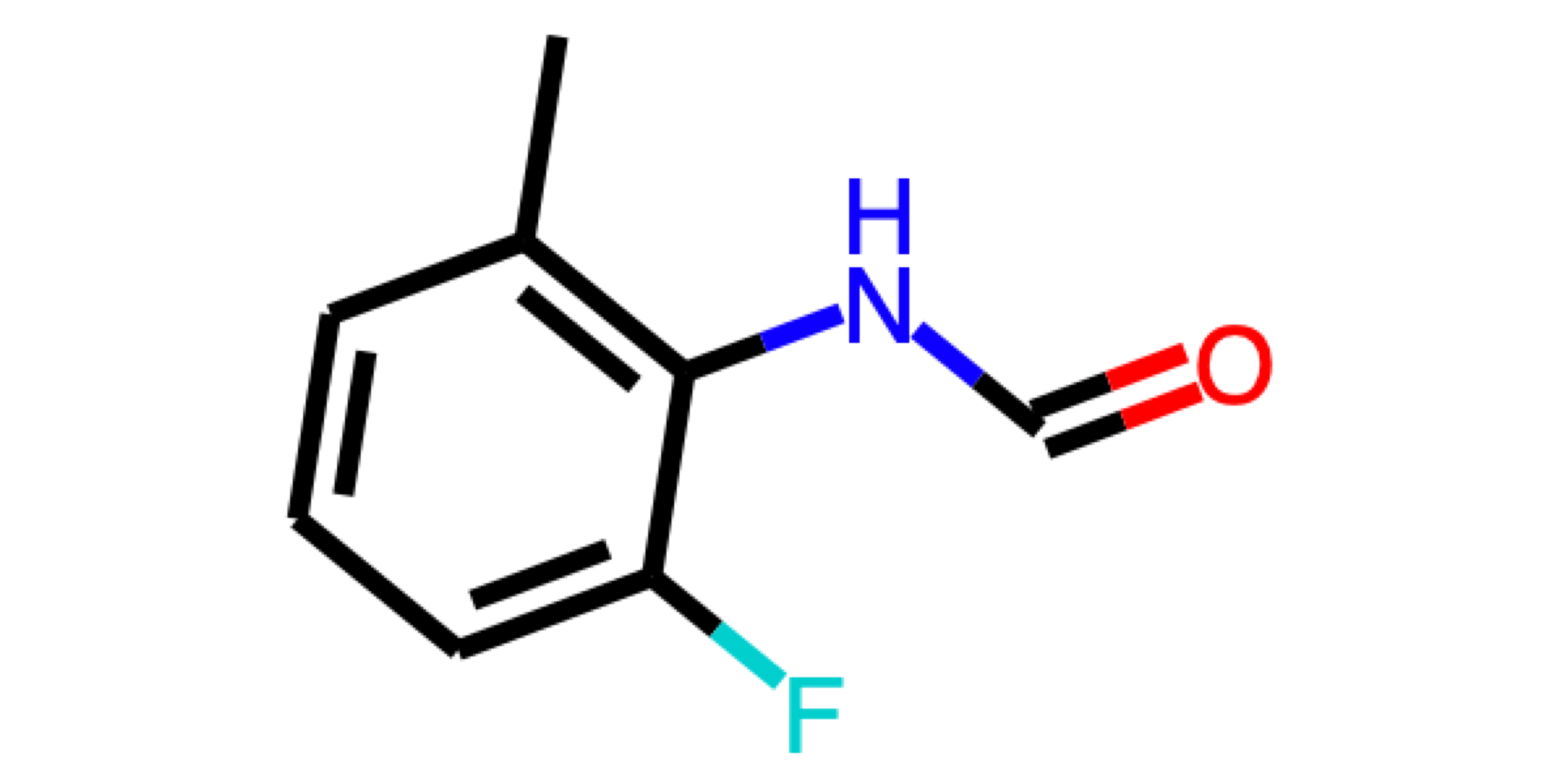}
    \caption{}
    \label{subfig:ranolazine_frag1}
    \end{subfigure}
    \begin{subfigure}{.3\textwidth}
    \centering
    \includegraphics[width=1.0\textwidth]{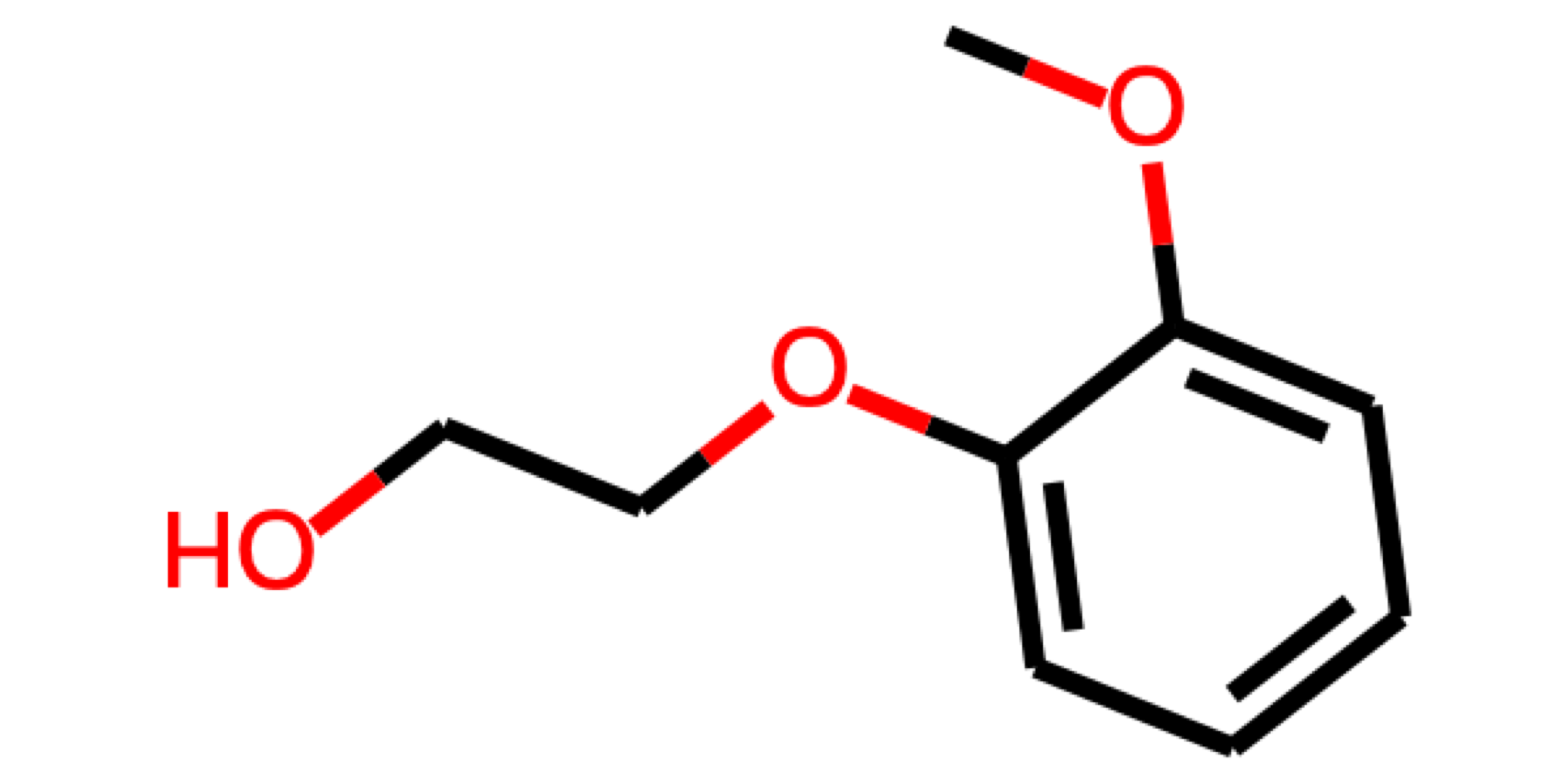}
    \caption{}
    \label{subfig:ranolazine_frag2}
    \end{subfigure}
    \caption{(\subref{subfig:ranolazine_passed}) The main fragments in the passed molecule of Ranolazine MPO are marked. Two main fragments exist in reality: (\subref{subfig:ranolazine_frag1}) N-(2-fluoro-6-methylphenyl) formamide, (\subref{subfig:ranolazine_frag2})
    2-(2-methoxyphenoxy) ethanol.}
    \label{fig:ranolazine_passed_whole}
    \vspace{-.15in}
\end{figure}

In this section, we further provide descriptions on the molecules generated from GEGL for the tasks of GuacaMol benchmark, illustrated in Figure \ref{fig:guacamol_duplicate}.

\textbf{Molecules passed from the post-hoc filtering.}
We first report some of the discovered molecules that pass the post-hoc filtering procedure and have realistic yet novel structures. For example, one may argue that the passed molecule for the Ranolazine MPO (in Figure \ref{fig:guacamol_duplicate}) is chemically realistic; its main fragments of N-(2-fluoro-6-methylphenyl) formamide,  (Figure \ref{subfig:ranolazine_frag1}) and  2-(2-methoxyphenoxy) ethanol (Figure \ref{subfig:ranolazine_frag2}) actually exist in reality. We have also verified the molecule to be novel from the chemical databases such as Scifinder \citep{gabrielson2018scifinder} and Drugbank \citep{wishart2018drugbank}. We provide the detailed illustration for the discovered molecule in Figure \ref{fig:ranolazine_passed_whole}.

% For example, the passed molecule for the Albuterol similarity has a similar structure with \textit{salbutamol}, which is used for relaxing bronchospasm. It means that we can expect the passed molecule is chemically realistic. Moreover, there does not exist the passed molecule in the chemical database, i.e., Scifinder and Drugbank. It means that the passed molecule is new one. 

\textbf{Molecules rejected from the post-hoc filtering.}
In Figure \ref{fig:guacamol_duplicate}, we illustrate an example of rejected molecules for the Albuterol similarity, Ranolazine MPO, Sitagliptin MPO, and Zaleplon MPO, respectively. They are filtered out as they are chemically reactive with high probability. Specifically, the molecule from the Albuterol similarity was rejected due to containing a SMILES of \texttt{SH}, i.e., \textit{I5 Thiols}. The molecule from the Ranolazine MPO was due to \texttt{C=CC=C}\footnote{Note that \texttt{C=CC=C} in a ring is stable, but \texttt{C=CC=C} not in a ring is reactive, i.e., the left side of the molecule.}, and that from the Sitagliptin MPO was due to \texttt{NNC=O}, i.e., \textit{acylhydrazide}. Functional groups like \texttt{SH}, \texttt{C=CC=C}, and \texttt{NNC=O} are known to be reactive with high probability, hence they are rejected from the filtering procedure.
%\section{Experiments with standard deviation}
%\input{trash/3_guacamol}
%\input{trash/4_guacamol_filtered}

%\input{trash/3_guacamol}
%\input{trash/4_guacamol_filtered}

\end{document}